\documentstyle[aps,prb]{revtex}
\begin{document}
\draft
\author{M. Valiev and G. W. Fernando}
\title{Generalized Kohn-Sham Density-Functional Theory via Effective Action
Formalism.}
\address{Physics Department, University of Connecticut,\\
Storrs, Connecticut, 06269, USA}
\date{February 26, 1997 }
\maketitle

\begin{abstract}
We present a rigorous formulation of generalized Kohn-Sham
density-functional theory. This provides a straightforward Kohn-Sham
description of many-body systems based not only on particle-density but also
on any other observable. We illustrate the formalism for the case of a
particle-density based description of a nonrelativistic many-electron
system. We obtain a simple diagrammatic expansion of the
exchange-correlation functional in terms of Kohn-Sham single-particle
orbitals and energies; develop systematic Kohn-Sham formulation for one
electron propagators and many-body excitation energies. This work is ideally
suited for practical applications and provides a rigorous basis for a
systematic development of the existing body of first-principles calculations
in a controllable fashion.
\end{abstract}


\section{Introduction}

Density-functional theory \cite{HK} allows one to study the ground state
properties of the many-body system in terms of the expectation value of the
particle-density operator. In principle, it offers the possibility of
finding the ground state energy $E_g$ by minimizing the energy functional
that depends on the density only:

\begin{equation}
E_g=\min\limits_nE\left[ n\right] .  \label{enfunc}
\end{equation}
Similar theories can be formulated in terms of the expectation values of a
spin-density operator or current-density operator, known as spin-density\cite
{SDFT} and current-density functional theory.\cite{CDFT} In general, one can
imagine a description of a many-body system in terms of the expectation
value of any other suitable operator. Such a general description can be
elegantly presented via the effective action formalism,\cite{fukuda,fukuda1}
thus leading to a {\it generalized} density-functional theory -- a theory
that allows a description of many-body systems in terms of the expectation
value of any suitable operator.

Proving the existence of such a theory is not a trivial matter. In fact, the
conventional density-functional theory relies heavily on the theorem of
Hohenberg and Kohn,\cite{HK} which shows that there exists a {\it unique}
description of a many-body system in terms of the expectation value of the
particle-density operator. Finite-temperature extension of this theorem was
given by Mermin.\cite{Mermin} Similar results have been obtained for
spin-density and current-density-functional theory.\cite{SDFT,CDFT} Clearly,
it is absolutely important to establish the corresponding existence theorems
when constructing a generalized density-functional theory. In the framework
of the effective action formalism, the proof of existence was given only in
a diagrammatic sense and was tightly tailored to the particular features of
the particle-density based description of a nonrelativistic many-electron
system.\cite{fukuda} In this work we present a different resolution of this
important issue. Our formulation is valid for a general case and does not
rely on any perturbative expansions. Not only does this offer an alternative
proof of the previous results for density-, spin-density-,
current-density-functional theories, but it also provides a rigorous
foundation for generalized density-functional theory.

However important the formal question of existence is, it is of little help
for the actual construction of the required density or other functionals.
For example, to make any practical use of the conventional
density-functional theory, an explicit (perhaps approximate) expression for
the energy functional $\left( \ref{enfunc}\right) $ is necessary. An
important contribution here was made by Kohn and Sham.\cite{KS} They
proposed a certain decomposition of the energy functional, which for a
typical nonrelativistic many-fermion system, 
\begin{eqnarray}
\hat H &=&\int d{\bf x}\hat \psi ^{\dagger }({\bf x})\left( -\frac 1{2m}%
\nabla ^2+\upsilon _{ion}({\bf x})\right) \hat \psi ({\bf x})  \nonumber \\
&&\ +\frac{e^2}2\int \int \frac{\hat \psi ^{\dagger }({\bf x})\hat \psi
^{\dagger }({\bf y})\hat \psi ({\bf y})\hat \psi ({\bf x})}{|{\bf x}-{\bf y}|%
}d{\bf x}d{\bf y},  \label{Ho}
\end{eqnarray}
takes the form

\begin{eqnarray}
E\left[ n\right] &=&T_{s}\left[ n\right] +\int \upsilon _{ion}({\bf x}%
)n\left( {\bf x}\right) d{\bf x}  \nonumber \\
&&{\bf +}\frac{e^{2}}{2}\int \int \frac{n({\bf x})n({\bf y})}{|{\bf x}-{\bf y%
}|}d{\bf x}d{\bf y}+E_{xc}\left[ n\right] .  \label{KSfunctional}
\end{eqnarray}
Here $T_{s}\left[ n\right] $ is the kinetic energy of an auxiliary system of
noninteracting fermions that yields the ground state density $n\left( {\bf x}%
\right) $, and $E_{xc}\left[ n\right] $ is the exchange-correlation energy.
Once the approximation for $E_{xc}\left[ n\right] $ has been decided, the
minimization of the functional $\left( \ref{KSfunctional}\right) $ leads to
the familiar Kohn-Sham single-particle equations.\cite{HKS} This approach
represents the so-called Kohn-Sham density-functional theory.

A natural question now arises: is there an analog of Kohn-Sham
density-functional theory in the effective action formalism? In the original
work of Fukuda, Kotani, Suzuki and Yokojima\cite{fukuda} the relationship
between the two methods was not established. Understanding the place of
Kohn-Sham theory in the effective action formalism is important for a number
of reasons. A vast majority of first-principles calculations are based on
the Kohn-Sham method.\cite{DFTbook1,DFTbook2} The remarkable success of
these calculations points to the fact that this seemingly ad hoc
decomposition provides a very good approximation of the energy functional.
As we show in this work, the connection between the effective action
formalism and the Kohn-Sham method can be rigorously established via the
inversion method.\cite{okumura} The realization of this fact immediately
leads to a generalized Kohn-Sham theory.

A distinct feature of the effective action formulation of generalized
Kohn-Sham theory is that it provides a systematic way of calculating the
required functionals. For example, in case of particle-density based
description of nonrelativistic many-electron systems, it leads to a set of
simple diagrammatic rules for constructing the exchange-correlation
functional entirely in terms of Kohn-Sham derived quantities. We construct
the first few orders of the exchange-correlation functional, comment on the
local density approximation, and discuss our results as compared to other
methods.\cite{DFTbook1,sham,OEP}

Applications of the presented formalism is not restricted to the
exchange-correlation functional only. We have already demonstrated that this
method is capable of establishing rigorous Kohn-Sham density-functional
formulation of one-electron propagators.\cite{valiev} We briefly discuss the
main results of that work and compare it with the existing strategies for
construction of one-electron propagators.\cite{SK,GW,GW1} Lastly, we analyze
the excitation energies within the effective action formalism and comment on
the relationship with similar results\cite{peter,excitation1} obtained via
time-dependent density-functional theory.\cite{GH}

\section{Effective action functional}

The effective action formalism in the context of density-functional theory
was discussed by Fukuda, Kotani, Suzuki and Yokojima.\cite{fukuda} Here we
describe the main features of this method and prove the generalized
existence theorems. We start by defining the functional $W\left[ J\right] $
as$,$ 
\[
e^{-\beta W\left[ J\right] }=\text{Tr}\left( e^{-\beta \left( \hat{H}%
+J\left( 1\right) \hat{Q}\left( 1\right) \right) }\right) . 
\]
Here $\hat{H}$ denotes the Hamiltonian of the system under consideration.
Parameter $\beta $ can be identified with inverse temperature. $\hat{Q}$ is
the operator whose expectation value will serve as a main variable of the
theory, and $J$ is the external source coupled to it. Both $\hat{H}$ and $%
\hat{Q}$ are assumed to be time-independent. Summation over repeated indexes
is assumed, and the notation $J\left( 1\right) \hat{Q}\left( 1\right) $
embodies all the necessary summations and integrations. For example, to
formulate the theory in terms of the expectation value of the
particle-density operator $\hat{n}\left( {\bf x}\right) $, we choose $\hat{Q}%
=\hat{n}\left( {\bf x}\right) $ and 
\[
J\left( 1\right) \hat{Q}\left( 1\right) \equiv \int d{\bf x}J\left( {\bf x}%
\right) \hat{n}\left( {\bf x}\right) . 
\]
Thermal expectation value of $\hat{Q},$%
\[
Q\left( 1\right) =\frac{\text{Tr}\left( \hat{Q}\left( 1\right) e^{-\beta
\left( \hat{H}+J\left( 1^{\prime }\right) \hat{Q}\left( 1^{\prime }\right)
\right) }\right) }{\text{Tr}\left( e^{-\beta \left( \hat{H}+J\left(
1^{\prime }\right) \hat{Q}\left( 1^{\prime }\right) \right) }\right) }, 
\]
can be written in terms of $W\left[ J\right] $ as, 
\begin{equation}
Q\left( 1\right) =\frac{\delta W\left[ J\right] }{\delta J\left( 1\right) }.
\label{Q=W/J}
\end{equation}
Denoting the set of allowable external sources as ${\cal J}$ and the set of
all generated expectation values of $Q$ as ${\cal Q}$, we can establish a
map ${\cal J}\rightarrow {\cal Q}.$ We assume that for a given element of $%
{\cal J}$ there corresponds only one element of ${\cal Q},$ in other words,
given the external source we can unambiguously establish the expectation
value that it generates. Whether the converse of this statement is true
remains to be proven. The following property of the functional $W\left[
J\right] $ is of fundamental importance.

{\bf Theorem 1}: {\it The functional }$W\left[ J\right] ${\it \ is strictly
concave, i.e. for any }$\alpha ${\it , }$0<\alpha <1,${\it \ and }$J\neq
J^{\prime }$

\[
W\left[ \alpha J+\left( 1-\alpha \right) J^{\prime }\right] >\alpha W\left[
J\right] +\left( 1-\alpha \right) W\left[ J^{\prime }\right] . 
\]
Based on strict concavity of $W\left[ J\right] $ one can prove the following
result.

{\bf Corollary 1:} {\it The map }${\cal J}\rightarrow {\cal Q}${\it \ is
one-to-one.}

The proofs of these statements are given in the appendix. Corollary 1
guarantees that the functional relationship between $J$ and $Q$ can be
inverted: 
\begin{equation}
\frac{\delta W\left[ J\right] }{\delta J\left( 1\right) }=Q\left( 1\right)
\Rightarrow \;J=J\left[ Q\right] .  \label{J=J[Q]}
\end{equation}
When $\hat Q=\hat n\left( {\bf x}\right) $, Corollary 1 represents an
alternative proof to the theorem of Mermin.\cite{Mermin}

The functional $W\left[ J\right] $ provides a description of the physical
system in terms of the external probe $J$. We, on the other hand, want the
description in terms of $Q$. The change of variables from $J$ to $Q$ can be
accomplished via a functional Legendre transformation.\cite{legendre1} This
leads to the definition of the effective action functional: 
\begin{equation}
\Gamma \left[ Q\right] =W\left[ J\right] -J\left( 1^{\prime }\right) Q\left(
1^{\prime }\right) .  \label{G[Q]}
\end{equation}
Here, $J$ is assumed to be a functional of $Q$ from Eqn $\left( \ref{J=J[Q]}%
\right) .$ The functional $\Gamma \left[ Q\right] $ possesses the following
important property.

{\bf Proposition 1: }{\it The effective action functional }$\Gamma \left[
Q\right] ${\it \ defined on the set }${\cal Q}${\it \ is strictly convex.}

Proof of this statement is given in the appendix. Differentiating $\left( 
\ref{G[Q]}\right) $ with respect to $Q$, we obtain 
\begin{equation}
\frac{\delta \Gamma \left[ Q\right] }{\delta Q\left( 1\right) }=-J\left(
1\right) .  \label{dG[Q]/dQ=-J}
\end{equation}
Since our original system is recovered when $J=0,$ we arrive at the
important variational principle: the functional $\Gamma \left[ Q\right] $
reaches a minimum at the exact expectation value of $\hat Q,$%
\[
\left( \frac{\delta \Gamma \left[ Q\right] }{\delta Q}\right) _{Q=Q_g}=0. 
\]
In the zero temperature limit, $\beta \rightarrow \infty $, $Q_g$ represents
the exact ground state expectation value of $\hat Q$ and $\Gamma \left[
Q_g\right] $ equals the exact ground state energy. Obviously, at finite
temperatures 
\[
\Gamma \left[ Q_g\right] =-\frac 1\beta \ln \text{Tr}\left( e^{-\beta \hat H%
}\right) , 
\]
\[
Q_g\left( 1\right) =\frac{\text{Tr}\left( \hat Q\left( 1\right) e^{-\beta 
\hat H}\right) }{\text{Tr}\left( e^{-\beta \hat H}\right) }. 
\]

The effective action formalism furnishes a rigorous formulation of
generalized density-functional theory. The existence of $\Gamma \left[
Q\right] $ is guaranteed by Corollary 1. When $\hat{Q}$ is a
particle-density operator, we obtain conventional density-functional theory;
when $\hat{Q}$ stands for a spin-density operator, we have
spin-density-functional theory;\cite{SDFT} when $\hat{Q}$ is a current
operator, we obtain current-density-functional theory,\cite{CDFT}, etc.

To implement this formally exact method, an approximation of the effective
action functional $\Gamma \left[ Q\right] $ is required. For a
particle-density based description of a nonrelativistic many-electron system
the effective action functional can be approximated via the auxiliary field
method.\cite{fukuda} However, this method does not lead\cite{fukuda} to
Kohn-Sham version of density-functional theory. The relationship between the
effective action formalism and Kohn-Sham density-functional theory can be
established via the inversion method.\cite{okumura}

\section{Generalized Kohn-Sham theory via the inversion method}

Consider the following general Hamiltonian 
\[
\hat H=\hat H_0+\lambda \hat H_{int} 
\]
which depends on the coupling constant $\lambda $ as a parameter. The same
is true for the effective action functional 
\[
\Gamma =\Gamma \left[ Q,\lambda \right] . 
\]
Clearly $Q,\lambda $ are to be considered as two independent variables.
Note, however, that this does not prevent the exact expectation value $Q_g$
from depending on $\lambda $: this dependence is fixed by the variational
principle 
\begin{equation}
\left( \frac{\delta \Gamma \left[ Q,\lambda \right] }{\delta Q}\right)
_{Q_g}=0.  \label{var}
\end{equation}
The functional $\Gamma \left[ Q,\lambda \right] $ is defined as, 
\begin{equation}
\Gamma \left[ Q,\lambda \right] =W\left[ J,\lambda \right] -J\left(
1^{\prime }\right) Q\left( 1^{\prime }\right) ,  \label{G[n]}
\end{equation}
where $J$ is functional of $Q$ and $\lambda $. This functional dependence is
provided by the equation 
\[
\frac{\delta W\left[ J,\lambda \right] }{\delta J\left( 1\right) }=Q\left(
1\right) . 
\]
The inversion method\cite{fukuda1,okumura} proceeds by expanding all the
quantities in Eqn $\left( \ref{G[n]}\right) $ in terms of $\lambda ;$ 
\[
J\left[ Q,\lambda \right] =J_0\left[ Q\right] +\lambda J_1\left[ Q\right]
+\lambda ^2J_2\left[ Q\right] +\ldots , 
\]
\begin{equation}
W\left[ J,\lambda \right] =W_0\left[ J\right] +\lambda W_1\left[ J\right]
+\lambda ^2W_2\left[ J\right] +\ldots ,  \label{W[n]=sum}
\end{equation}
\[
\Gamma \left[ Q,\lambda \right] =\Gamma _0\left[ Q\right] +\lambda \Gamma
_1\left[ Q\right] +\lambda ^2\Gamma _2\left[ Q\right] +\ldots . 
\]
Comparison of the two sides in Eqn $\left( \ref{G[n]}\right) $ for different
orders of $\lambda $, 
\begin{eqnarray}
\sum \lambda ^i\Gamma _i\left[ Q\right] &=&\sum \lambda ^iW_i\left[ \sum
\lambda ^kJ_k\left[ Q\right] \right]  \nonumber \\
&&\ \ -\sum \lambda ^iJ_i\left( 1\right) Q\left( 1\right) ,
\end{eqnarray}
leads to the expression for $\Gamma _l\left[ Q\right] ,$ 
\begin{eqnarray*}
\Gamma _l\left[ Q\right] =W_l\left[ J_0\right] +\sum_{k=1}^l\frac{\delta
W_{l-k}\left[ J_0\right] }{\delta J_0\left( 1\right) } &&J_k\left( 1\right)
-J_l\left( 1\right) Q\left( 1\right) \\
+\sum_{m=2}^l\frac 1{m!}\sum_{k_1,\ldots ,k_m\geq 1}^{k_1+\ldots +k_m\leq l}
&&\frac{\delta ^mW_{l-\left( k_1+\ldots +k_m\right) }\left[ J_0\right] }{%
\delta J_0\left( 1\right) \ldots \delta J_0\left( m\right) }\times \\
&&\ \ \times J_{k_1}\left( 1\right) \cdots J_{k_m}\left( m\right)
\end{eqnarray*}
Functionals $\left\{ W_l\left[ J_0\right] \right\} $ and its derivatives are
assumed to be known;\ they can usually be obtained via standard many-body
perturbation techniques (specific examples will be given in the next
section). Since $Q$ and $\lambda $ are considered to be independent, it
follows from Eqn $\left( \ref{dG[Q]/dQ=-J}\right) $ that functionals $%
\left\{ J_k\left[ Q\right] \right\} $ can be obtained using$,$%
\begin{equation}
\frac{\delta \Gamma _k\left[ Q\right] }{\delta Q\left( 1\right) }=-J_k\left(
1\right) .  \label{dGl[Q]/dQ=-Jl}
\end{equation}
Consider the zeroth order term, 
\begin{equation}
\Gamma _0\left[ Q\right] =W_0\left[ J_0\right] -J_0\left( 1\right) Q\left(
1\right) .  \label{G0}
\end{equation}
Using Eqn $\left( \ref{dGl[Q]/dQ=-Jl}\right) $%
\[
-J_0\left( 1\right) =\frac{\delta W_0\left[ J_0\right] }{\delta J_0\left(
1^{\prime }\right) }\frac{\delta J_0\left( 1^{\prime }\right) }{\delta
Q\left( 1\right) }-J_0\left( 1\right) -Q\left( 1^{\prime }\right) \frac{%
\delta J_0\left( 1^{\prime }\right) }{\delta Q\left( 1\right) } 
\]
\[
\Rightarrow \left( \frac{\delta W_0\left[ J_0\right] }{\delta J_0\left(
1^{\prime }\right) }-Q\left( 1^{\prime }\right) \right) \frac{\delta
J_0\left( 1^{\prime }\right) }{\delta Q\left( 1\right) }=0. 
\]
Strict convexity of $\Gamma _0\left[ Q\right] $ (see Proposition 1)
prohibits $\left( \delta J_0\left( 1^{\prime }\right) /\delta Q\left(
1\right) \right) $ from having zero eigenvalues. Thus we obtain that $J_0$
obeys the equation: 
\begin{equation}
Q\left( 1\right) =\frac{\delta W_0\left[ J_0\right] }{\delta J_0\left(
1\right) }.  \label{J0}
\end{equation}
Hence $J_0$ is determined as a potential which generates the expectation
value $Q$ in the {\it noninteracting} $\left( \lambda =0\right) $ system.
Notice that the same exact notion appears in Kohn-Sham formalism.\cite{HKS}
We refer to this noninteracting system as Kohn-Sham (KS) system and $J_0$ as
Kohn-Sham potential. Eqn $\left( \ref{J0}\right) $ allows one to simplify
the expression for $\Gamma _l\left[ Q\right] $, which now becomes, 
\begin{eqnarray*}
\Gamma _l\left[ Q\right] =W_l\left[ J_0\right] -\delta _{l,0}J_0\left(
1\right) Q &&\left( 1\right) +\sum_{k=1}^{l-1}\frac{\delta W_{l-k}\left[
J_0\right] }{\delta J_0\left( 1\right) }J_k\left( 1\right) \\
+\sum_{m=2}^l\frac 1{m!}\sum_{k_1,\ldots ,k_m\geq 1}^{k_1+\ldots +k_m\leq l}
&&\frac{\delta ^mW_{l-\left( k_1+\ldots +k_m\right) }\left[ J_0\right] }{%
\delta J_0\left( 1\right) \ldots \delta J_0\left( m\right) }\times \\
&&\ \ \times J_{k_1}\left( 1\right) \cdots J_{k_m}\left( m\right) .
\end{eqnarray*}
The important message here is that the expression for $\Gamma _l\left[
Q\right] $ involves only lower order functionals ($l-1,l-2,\ldots ,0)$. Thus
starting with $J_0$ we can determine $\Gamma _1\left[ Q\right] $ as 
\begin{equation}
\Gamma _1\left[ Q\right] =W_1\left[ J_0\right] .  \label{G1}
\end{equation}
From $\Gamma _1\left[ Q\right] $ we can find $J_1$ as, 
\[
J_1\left( 1\right) =-\frac{\delta \Gamma _1\left[ Q\right] }{\delta Q\left(
1\right) }=-\frac{\delta \Gamma _1\left[ Q\right] }{\delta J_0\left(
1^{\prime }\right) }\frac{\delta J_0\left( 1^{\prime }\right) }{\delta
Q\left( 1\right) }, 
\]
or 
\begin{equation}
J_1\left( 1\right) ={\cal D}\left( 1,2\right) \frac{\delta W_1\left[
J_0\right] }{\delta J_0\left( 2\right) },  \label{J1[Q]}
\end{equation}
where the inverse propagator is defined as, 
\[
{\cal D}\left( 1{\bf ,}2\right) =-\frac{\delta J_0\left( 2\right) }{\delta
Q\left( 1\right) }=-\left( \frac{\delta ^2W_0\left[ J_0\right] }{\delta
J_0\left( 1\right) \delta J_0\left( 2\right) }\right) ^{-1}. 
\]
Once $J_1$ is known, we can find $\Gamma _2\left[ Q\right] :$%
\begin{eqnarray*}
\Gamma _2\left[ Q\right] &=&W_2\left[ J_0\right] +\frac{\delta W_1\left[
J_0\right] }{\delta J_0\left( 1\right) }J_1\left( 1\right) \\
&&\ \ \ \ \ \ \ \ +\frac 12\frac{\delta ^2W_0\left[ J_0\right] }{\delta
J_0\left( 1\right) \delta J_0\left( 2\right) }J_1\left( 1\right) J_1\left(
2\right) ,
\end{eqnarray*}
or using Eqn $\left( \ref{J1[Q]}\right) $%
\begin{equation}
\Gamma _2\left[ Q\right] =W_2\left[ J_0\right] +\frac 12\frac{\delta
W_1\left[ J_0\right] }{\delta J_0\left( 1\right) }{\cal D}\left( 1,2\right) 
\frac{\delta W_1\left[ J_0\right] }{\delta J_0\left( 2\right) }.  \label{G2}
\end{equation}
From $\Gamma _2\left[ Q\right] $ the expression for $J_2$ follows as, 
\begin{eqnarray}
J_2\left( 1\right) &=&{\cal D}\left( 1,2\right) \frac{\delta W_2\left[
J_0\right] }{\delta J_0\left( 2\right) }  \nonumber \\
&&\ \ {\bf +}{\cal D}\left( 1,2\right) \frac{\delta ^2W_1\left[ J_0\right] }{%
\delta J_0\left( 2\right) \delta J_0\left( 2^{\prime }\right) }J_1\left(
2^{\prime }\right)  \nonumber \\
&&\ \ \ {\bf +}\frac 12{\cal D}\left( 1,2\right) \frac{\delta ^3W_0\left[
J_0\right] }{\delta J_0\left( 2\right) \delta J_0\left( 3\right) \delta
J_0\left( 4\right) }J_1\left( 3\right) J_1\left( 4\right) {\bf .}
\label{J2[Q]}
\end{eqnarray}
This, in turn, leads to $\Gamma _3\left[ Q\right] $ and so on. In this
hierarchical fashion one can consistently determine $\Gamma \left[ Q,\lambda
\right] $ to any required order$;$%
\[
\Gamma _0\rightarrow J_0\rightarrow \Gamma _1\rightarrow J_1\rightarrow
\Gamma _2\rightarrow J_2\rightarrow \ldots . 
\]
The important point here is that all higher orders are completely determined
by the Kohn-Sham potential $J_0$. Let us now apply the variational principle
to our expansion of $\Gamma \left[ Q,\lambda \right] :$%
\[
\frac{\delta \Gamma \left[ Q,\lambda \right] }{\delta Q\left( 1\right) }=0. 
\]
Since 
\[
\frac{\delta \Gamma _0\left[ Q\right] }{\delta Q\left( 1\right) }=-J_0\left(
1\right) , 
\]
we have the following important result 
\begin{equation}
J_0\left( 1\right) =\frac{\delta \Gamma _{int}\left[ Q\right] }{\delta
Q\left( 1\right) }=-{\cal D}\left( 1,1^{\prime }\right) \frac{\delta \Gamma
_{int}\left[ Q\right] }{\delta J_0\left( 1^{\prime }\right) },
\label{genKSpotential}
\end{equation}
where 
\[
\Gamma _{int}\left[ Q\right] =\sum_{i=1}\lambda ^i\Gamma _i\left[ Q\right] . 
\]
Since at any order the effective action functional is completely determined
by Kohn-Sham potential $J_0,$ we arrive at generalized Kohn-Sham
self-consistent method:

\begin{enumerate}
\item  Choose the approximation for $\Gamma _{int}\left[ Q\right] $ (one
obvious choice is to truncate the expansion at some order).

\item  Start with some reasonable guess for the Kohn-Sham potential $J_0.$

\item  Calculate $\Gamma _{int}\left[ Q\right] .$

\item  Determine new Kohn-Sham potential $J_0$ via Eqn $\left( \ref
{genKSpotential}\right) .$

\item  Repeat from step 3 until self-consistency is achieved.
\end{enumerate}

The formalism described above can be applied to any general case and
provides a rigorous basis for generalized Kohn-Sham theory. Practical
implementation of the self-consistent procedure obviously depends on the
particular Hamiltonian under consideration and the choice of the operator $%
\hat Q$. In the next section we apply this method to the case of a
particle-density based description of nonrelativistic many-electron system.

\section{Kohn-Sham density-functional theory.}

\subsection{Derivation of Kohn-Sham decomposition}

Let us consider a typical nonrelativistic many-electron system described by
the Hamiltonian $\left( \ref{Ho}\right) $ and develop the description in
terms of the particle-density operator: 
\[
\hat Q\left( 1\right) =\hat \psi ^{\dagger }({\bf x})\hat \psi ({\bf x}%
)\equiv \hat n({\bf x}). 
\]
For convenience, the spin degrees of freedom are suppressed here (those can
be easily recovered if necessary). The role of the coupling constant $%
\lambda $ is played by $e^2$. We now evaluate the effective action
functional $\Gamma \left[ n\right] $ using the inversion method\cite{okumura}
described in the previous section. Coupling constant expansion of the
functional $W\left[ J_0,e^2\right] ,$%
\[
W\left[ J_0,e^2\right] =W_0\left[ J_0\right] +e^2W_1\left[ J_0\right]
+e^4W_2\left[ J_0\right] +\ldots , 
\]
can be conveniently generated using path integral representation\cite{negele}
\begin{equation}
e^{-\beta W\left[ J\right] }=\int D\psi ^{\dagger }D\psi \;e^{-S\left[ \psi
^{\dagger },\psi \right] -\int J\left( {\bf x}\right) \psi ^{\dagger }\left(
x\right) \psi \left( x\right) dx}.  \label{genfunc}
\end{equation}
Here 
\begin{eqnarray}
S &&\left[ \psi ^{\dagger },\psi \right] =\int dx\;\psi ^{\dagger }(x)\left[ 
\frac \partial {\partial \tau }-\frac{\nabla ^2}{2m}+\upsilon _{ion}({\bf x}%
)\right] \psi (x)  \nonumber \\
&&+\frac{e^2}2\int \int dxdx^{\prime }\psi ^{\dagger }(x)\psi ^{\dagger
}(x^{\prime })u(x-x^{\prime })\psi (x^{\prime })\psi (x),  \label{action}
\end{eqnarray}
\[
u(x-x^{\prime })=\frac{\delta \left( \tau -\tau ^{\prime }\right) }{|{\bf x}-%
{\bf x}^{\prime }|}, 
\]
\[
\int dx\equiv \int_0^\beta d\tau \int d{\bf x.} 
\]
and $\psi ^{\dagger },\psi $ denote Grassmann fields.\cite{negele}

The zeroth order term is given by, 
\[
W_0\left[ J_0\right] =-\frac 1\beta \sum_i\ln \left( 1+e^{-\beta \varepsilon
_i}\right) , 
\]
where $\varepsilon _i$'s denote single-particle energies of Kohn-Sham
noninteracting system:

\[
\left( -\frac{\nabla ^2}{2m}+\upsilon _{ion}({\bf x})+J_0\left( {\bf x}%
\right) \right) \varphi _i\left( {\bf x}\right) =\varepsilon _i\varphi
_i\left( {\bf x}\right) ,
\]
\[
n({\bf x})=\sum_in_i\left| \varphi _i\left( {\bf x}\right) \right| ^2,%
\hspace{0.5cm}n_i=\left( e^{\beta \varepsilon _i}+1\right) ^{-1}.
\]
First order term is given by, 
\begin{equation}
W_{1}\left[ J_{0}\right] =\setlength{\unitlength}{1cm}\frac{1}{2\beta }{}\;%
\begin{picture}(1,1)(0,0.4) \put(0.5,0.5){\circle{1}}
\put(0.01,0.5){\vector(0,-1){0}} \end{picture}{}%
\begin{picture}(0.5,1)(0,0.4) 
\multiput(0,0.5)(0.082,0){7}{\circle*{0.05}}
\end{picture}{}%
\begin{picture}(1,1)(0,0.4) \put(0.5,0.5){\circle{1}}
\put(0.99,0.5){\vector(0,-1){0.0}} \end{picture}-\frac{1}{2\beta }{}\;%
\begin{picture}(1,1)(0,0.4) \put(0.5,0.5){\circle{1}}
\put(0.01,0.5){\vector(0,-1){0}} \put(0.99,0.5){\vector(0,1){0}}
\multiput(0.5,0)(0,0.083){13}{\circle*{0.05}}
\end{picture}
\label{W1}
\end{equation}
Here solid lines denote Matsubara Green's function of Kohn-Sham
noninteracting system 
\[
{\cal G}_0({\bf x}\tau ,{\bf x}^{\prime }\tau ^{\prime })=\left\{ 
\begin{array}{l}
\tau >\tau ^{\prime },\sum {\small \varphi }_i\left( {\bf x}\right) {\small %
\varphi }_i^{*}\left( {\bf x}^{\prime }\right) {\small e}^{-\varepsilon
_i\left( \tau -\tau ^{\prime }\right) }\left( n_i-1\right)  \\ 
\tau \leq \tau ^{\prime },\sum {\small \varphi }_i\left( {\bf x}\right) 
{\small \varphi }_i^{*}\left( {\bf x}^{\prime }\right) {\small e}%
^{-\varepsilon _i\left( \tau -\tau ^{\prime }\right) }n_i
\end{array}
\right. 
\]
and dotted line stands for the Coulomb interaction $u\left( x-x^{\prime
}\right) $. The explicit expression for $W_1\left[ J_0\right] $ is given by, 
\begin{eqnarray}
W_1\left[ J_0\right]  &=&\frac 1{2\beta }\int \int {\cal G}_0(x,x)u\left(
x-x^{\prime }\right) {\cal G}_0(x^{\prime },x^{\prime })dxdx^{\prime } 
\nonumber \\
&&\ -\frac 1{2\beta }\int \int {\cal G}_0(x,x^{\prime })u\left( x-x^{\prime
}\right) {\cal G}_0(x^{\prime },x)dxdx^{\prime }.
\end{eqnarray}
The second order term is 
\begin{eqnarray}
W_{2}\left[ J_{0}\right]  &=&\setlength{\unitlength}{1cm}\frac{1}{4\beta }\;%
\begin{picture}(1,1)(0,0.4)
\put(0.5,0.5){\circle{1}}
\multiput(0.15,0.15)(0.05,0.05){15}{\circle*{0.05}}
\multiput(0.85,0.15)(-0.05,0.05){15}{\circle*{0.05}}
\put(0.01,0.5){\vector(0,-1){0}} \put(0.99,0.5){\vector(0,1){0}}
\put(0.5,1){\vector(-1,0){0}} \put(0.5,0){\vector(1,0){0}} \end{picture}-%
\frac{1}{4\beta }\;%
\begin{picture}(1,1)(0,0.4) \put(0.5,0.5){\circle{1}} 
\multiput(0.85,0.85)(0.08,0){11}{\circle*{0.05}} 
\multiput(0.85,0.15)(0.08,0){11}{\circle*{0.05}} \put(0.01,0.5){\vector(0,-1){0}} \put(0.99,0.5){\vector(0,1){0}}\end{picture}%
\begin{picture}(0.5,1)(0,0.4) \end{picture}{}\begin{picture}(1,1)(0,0.4)
\put(0.5,0.5){\circle{1}} \put(0.01,0.5){\vector(0,-1){0}}
\put(0.99,0.5){\vector(0,1){0}}\end{picture} \\
&&+\setlength{\unitlength}{1cm}\frac{1}{2\beta }\;%
\begin{picture}(1,1)(0,0.4) \put(0.5,0.5){\circle{1}} 
\multiput(0.16,0.15)(0,0.07){11}{\circle*{0.05}} 
\multiput(0.84,0.15)(0,0.07){11}{\circle*{0.05}} \put(0.01,0.5){\vector(0,-1){0}} \put(0.99,0.5){\vector(0,1){0}}\put(0.5,1){\vector(-1,0){0}} \put(0.5,0){\vector(1,0){0}}\end{picture}%
-\frac{1}{\beta }\;\begin{picture}(1,1)(0,0.4) \put(0.5,0.5){\circle{1}}
\put(0.01,0.5){\vector(0,-1){0}} \end{picture}{}%
\begin{picture}(0.5,1)(0,0.4) 
\multiput(0,0.5)(0.082,0){7}{\circle*{0.05}} \end{picture}{}%
\begin{picture}(1,1)(0,0.4) \put(0.5,0.5){\circle{1}} 
\multiput(0.5,0)(0,0.083){13}{\circle*{0.05}} \put(0.15,0.15){\vector(-1,1){0}} \put(0.15,0.85){\vector(1,1){0}} \put(0.99,0.5){\vector(0,-1){0}} \end{picture}
\\
&&+\setlength{\unitlength}{1cm}\frac{1}{2\beta }\;%
\begin{picture}(1,1)(0,0.4) \put(0.5,0.5){\circle{1}}
\put(0.01,0.5){\vector(0,-1){0}} \end{picture}{}%
\begin{picture}(0.5,1)(0,0.4) \multiput(0,0.5)(0.082,0){7}{\circle*{0.05}}
\end{picture}{}\begin{picture}(1,1)(0,0.4) \put(0.5,0.5){\circle{1}}
\put(0.5,1){\vector(-1,0){0}} \put(0.5,0){\vector(1,0){0}}\end{picture}%
\begin{picture}(0.5,1)(0,0.4) \multiput(0,0.5)(0.082,0){7}{\circle*{0.05}}
\end{picture}\begin{picture}(1,1)(0,0.4) \put(0.5,0.5){\circle{1}}
\put(0.99,0.5){\vector(0,-1){0}}\end{picture}  
\label{W2}
\end{eqnarray}
Similarly, one can generate higher order terms. Let us now calculate the
first few orders of the effective action functional. Using Eqn $\left( \ref
{G0}\right) ,$ the zeroth order correction is given by, 
\[
\Gamma _0\left[ n\right] =-\frac 1\beta \sum_i\ln \left( 1+e^{-\beta
\varepsilon _i}\right) -\int J_0\left( {\bf x}\right) n\left( {\bf x}\right)
d{\bf x.}
\]
In the zero temperature limit, $\beta \rightarrow \infty ,$ this transforms
to 
\[
\Gamma _0\left[ n\right] =\sum_{i=1}^N\varepsilon _i-\int J_0\left( {\bf x}%
\right) n\left( {\bf x}\right) d{\bf x,}
\]
or 
\[
\Gamma _0\left[ n\right] =T_0\left[ n\right] +\int \upsilon _{ion}({\bf x}%
)n\left( {\bf x}\right) d{\bf x.}
\]
Hence at zeroth order, the effective action functional is given by the sum
of kinetic energy $T_0\left[ n\right] $ and ion-potential energy of the {\it %
Kohn-Sham noninteracting system}. From Eqn $\left( \ref{G1}\right) ,$ the
first order correction is given by, 
\[
\Gamma _1\left[ n\right] =W_1\left[ J_0\right] .
\]
The expression for $W_1\left[ J_0\right] $ has been presented earlier (see
Eqn $\left( \ref{W1}\right) $)$.$ In the zero temperature limit 
\begin{eqnarray*}
\Gamma _1\left[ n\right]  &=&\frac 12\int \int \frac{n\left( {\bf x}\right)
n\left( {\bf x}^{\prime }\right) }{\left| {\bf x}-{\bf x}^{\prime }\right| }d%
{\bf x}d{\bf x}^{\prime } \\
&&\ \ \ \ -\frac 12\int \int \frac{n\left( {\bf x,x}^{\prime }\right)
n\left( {\bf x}^{\prime },{\bf x}\right) }{\left| {\bf x}-{\bf x}^{\prime
}\right| }d{\bf x}d{\bf x}^{\prime },
\end{eqnarray*}
where $n\left( {\bf x,x}^{\prime }\right) =\sum \varphi _i\left( {\bf x}%
\right) \varphi _i^{*}\left( {\bf x}^{\prime }\right) .$ Here, the first
term represents the classical Hartree energy and the second term represents
the exchange energy. Both are evaluated with respect to Kohn-Sham
noninteracting system.

Postponing the evaluation of the second order term until the next section,
let us summarize the results we have obtained. It is clear that in the zero
temperature limit the expansion for the effective action functional, 
\begin{eqnarray}
\Gamma \left[ n\right] &=&T_{0}\left[ n\right] +\int \upsilon _{ion}({\bf x}%
)n\left( {\bf x}\right) d{\bf x}  \nonumber \\
&&\ \ {\bf +}\frac{e^{2}}{2}\int \int \frac{n\left( {\bf x}\right) n\left( 
{\bf x}^{\prime }\right) }{\left| {\bf x}-{\bf x}^{\prime }\right| }d{\bf x}d%
{\bf x}^{\prime }  \nonumber \\
&\ \ &-\frac{e^{2}}{2}\int \int \frac{n\left( {\bf x,x}^{\prime }\right)
n\left( {\bf x}^{\prime },{\bf x}\right) }{\left| {\bf x}-{\bf x}^{\prime
}\right| }d{\bf x}d{\bf x}^{\prime }  \nonumber \\
&&+\sum_{i=2}e^{2i}\Gamma _{i}\left[ n\right] ,  \label{KSfunctional1}
\end{eqnarray}
coincides with the decomposition proposed by Kohn and Sham.\cite{KS}
Therefore the inversion method of evaluating the effective action functional%
\cite{okumura} naturally leads to the Kohn-Sham density-functional theory.
Application of the variational principle (see Eqn $\left( \ref{var}\right) $
) to the expansion $\left( \ref{KSfunctional1}\right) $ yields the
well-known Kohn-Sham self-consistent procedure and the corresponding
single-particle equations. Comparison of Eqns $\left( \ref{KSfunctional}%
\right) $ and $\left( \ref{KSfunctional1}\right) $ immediately provides an
expression for the exchange-correlation functional. These topics are
discussed in detail in the following sections.

To conclude this section, we would like to note that alternatively, the
effective action functional can be also evaluated via the auxiliary field
method\cite{fukuda} However, in this method the Kohn-Sham decomposition has
to be artificially imposed to allow the study of the exchange-correlation
functional \cite{valiev1} or Kohn-Sham density-functional theory in general.

\subsection{Construction of the exchange-correlation functional}

The success of first-principles calculations based on Kohn-Sham
density-functional theory depends on the accuracy of the approximations to
the exchange-correlation functional. The analysis of Kohn-Sham theory, or $%
E_{xc}$ in particular, via standard many-body perturbation theory\cite
{DFTbook1,DFTbook2} was always a challenging task, for there was no explicit
connection between the two methods. The advantage of the effective action
formalism is that it is a rigorous many-body approach specifically designed
for a density-based description of many-body systems. This formalism
provides a natural definition of the exchange-correlation functional as, 
\[
E_{xc}\left[ n\right] =-\frac{1}{2\beta }\;\setlength{\unitlength}{1cm}%
\begin{picture}(1,1)(0,0.4) \put(0.5,0.5){\circle{1}}
\put(0.01,0.5){\vector(0,-1){0}} \put(0.99,0.5){\vector(0,1){0}}
\multiput(0.5,0)(0,0.083){13}{\circle*{0.05}}
\end{picture}
+\sum_{i=2}e^{2i}\Gamma _{i}\left[ n\right] .
\]
This expression involves only Kohn-Sham based quantities and is especially
suitable for practical applications. Simple diagrammatic rules for
evaluating higher order terms in the expansion of $\Gamma \left[ n\right] $
are readily available.\cite{fukuda1,okumura} This, in turn leads to the
following set of rules for the calculation of the exchange-correlation
functional:

\begin{enumerate}
\item  Draw all connected diagrams made of Kohn-Sham propagators ${\cal G}%
_0\left( x,x^{\prime }\right) $ and Coulomb interaction lines $u\left(
x-x^{\prime }\right) $ with the corresponding weight factors.\cite{negele}

\item  Eliminate all the graphs that can be separated by cutting a single
Coulomb interaction line.

\item  For each two-particle reducible (2PR) graph (i.e. any graph that can
be separated by cutting two propagator lines) perform the following
procedure.\cite{fukuda1}

\begin{enumerate}
\item  Separate the graph by cutting 2PR propagators.

\item  For each of the two resulting graphs join two external propagators.

\item  Connect the two graphs via the inverse density propagator ${\cal D}%
\left( {\bf x},{\bf x}^{\prime }\right) .$

\item  Repeat the procedure until no new graph is produced.

\item  Sum up all the resulting graphs including the original graph.
\end{enumerate}
\end{enumerate}

The inverse density propagator ${\cal D}\left( {\bf x},{\bf x}^{\prime
}\right) $ is given by, 
\[
{\cal D}\left( {\bf x},{\bf x}^{\prime }\right) =-\left[ \int_{0}^{\beta }%
{\cal G}_{0}\left( {\bf x}\tau ,{\bf x}^{\prime }\tau ^{\prime }\right) 
{\cal G}_{0}\left( {\bf x}^{\prime }\tau ^{\prime },{\bf x}\tau \right)
d\tau ^{\prime }\right] ^{-1}, 
\]
or in terms of Kohn-Sham orbitals 
\begin{equation}
{\cal D}\left( {\bf x},{\bf x}^{\prime }\right) =-\left[ \sum_{i\neq
j}\left( n_{i}-n_{j}\right) \frac{{\small \varphi }_{i}\left( {\bf x}\right) 
{\small \varphi }_{i}^{*}\left( {\bf x}^{\prime }\right) {\small \varphi }%
_{j}\left( {\bf x}^{\prime }\right) {\small \varphi }_{j}^{*}\left( {\bf x}%
\right) }{\varepsilon _{i}-\varepsilon _{j}}\right] ^{-1}.  \label{D(x,x')}
\end{equation}

Application of these rules for generation of the first-order correction to $%
E_{xc}\left[ n\right] $ is obvious and leads to the well-known Kohn-Sham
exchange functional:\cite{sham,OEP,valiev1,Levy} 
\[
E_{xc,1}=-\frac{1}{2\beta }\;\setlength{\unitlength}{1cm}%
\begin{picture}(1,1)(0,0.4) \put(0.5,0.5){\circle{1}}
\put(0.01,0.5){\vector(0,-1){0}} \put(0.99,0.5){\vector(0,1){0}}
\multiput(0.5,0)(0,0.083){13}{\circle*{0.05}}
\end{picture}
\]
Let us now consider the second order correction to $E_{xc}\left[ n\right] .$
Rule \# 1 leads to the already given expression for $W_2\left[ J_0\right] $
(see Eqn $\left( \ref{W2}\right) $). Application of Rule \# 2 leads to
elimination of the last two graphs, thus giving 
\[
\setlength{\unitlength}{1cm}\frac{1}{4\beta }\;\begin{picture}(1,1)(0,0.4)
\put(0.5,0.5){\circle{1}}
\multiput(0.15,0.15)(0.05,0.05){15}{\circle*{0.05}}
\multiput(0.85,0.15)(-0.05,0.05){15}{\circle*{0.05}}
\put(0.01,0.5){\vector(0,-1){0}} \put(0.99,0.5){\vector(0,1){0}}
\put(0.5,1){\vector(-1,0){0}} \put(0.5,0){\vector(1,0){0}} \end{picture}-%
\frac{1}{4\beta }\;\begin{picture}(1,1)(0,0.4) \put(0.5,0.5){\circle{1}}
\multiput(0.85,0.85)(0.08,0){11}{\circle*{0.05}}
\multiput(0.85,0.15)(0.08,0){11}{\circle*{0.05}}
\put(0.01,0.5){\vector(0,-1){0}} \put(0.99,0.5){\vector(0,1){0}}\end{picture}%
\begin{picture}(0.5,1)(0,0.4) \end{picture}{}\begin{picture}(1,1)(0,0.4)
\put(0.5,0.5){\circle{1}} \put(0.01,0.5){\vector(0,-1){0}}
\put(0.99,0.5){\vector(0,1){0}}\end{picture}+\frac{1}{2\beta }\;%
\begin{picture}(1,1)(0,0.4) \put(0.5,0.5){\circle{1}}
\multiput(0.16,0.15)(0,0.07){11}{\circle*{0.05}}
\multiput(0.84,0.15)(0,0.07){11}{\circle*{0.05}}
\put(0.01,0.5){\vector(0,-1){0}}
\put(0.99,0.5){\vector(0,1){0}}\put(0.5,1){\vector(-1,0){0}}
\put(0.5,0){\vector(1,0){0}}\end{picture}
\]
The third graph in the above expression is 2PR. According to Rule \# 3 it
transforms to: 
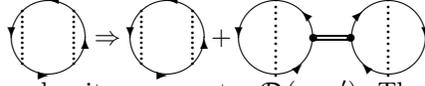
\[
\setlength{\unitlength}{1cm}\begin{picture}(1,1)(0,0.4)
\put(0.5,0.5){\circle{1}} \multiput(0.16,0.15)(0,0.07){11}{\circle*{0.05}}
\multiput(0.84,0.15)(0,0.07){11}{\circle*{0.05}}
\put(0.01,0.5){\vector(0,-1){0}}
\put(0.99,0.5){\vector(0,1){0}}\put(0.5,1){\vector(-1,0){0}}
\put(0.5,0){\vector(1,0){0}}\end{picture}\Rightarrow %
\begin{picture}(1,1)(0,0.4) \put(0.5,0.5){\circle{1}}
\multiput(0.16,0.15)(0,0.07){11}{\circle*{0.05}}
\multiput(0.84,0.15)(0,0.07){11}{\circle*{0.05}}
\put(0.01,0.5){\vector(0,-1){0}}
\put(0.99,0.5){\vector(0,1){0}}\put(0.5,1){\vector(-1,0){0}}
\put(0.5,0){\vector(1,0){0}}\end{picture}+%
\begin{picture}(1,1)(0,0.4) \put(0.5,0.5){\circle{1}} \multiput(0.5,0)(0,0.083){13}{\circle*{0.05}} 
\put(0.01,0.5){\vector(0,-1){0}}
\put(0.85,0.15){\vector(1,1){0}}
\put(0.85,0.85){\vector(-1,1){0}}
\end{picture} 
\begin{picture}(0.5,1)(0,0.4) \put(0,0.52){\line(1,0){0.5}}
\put(0,0.47){\line(1,0){0.5}} \put(0,0.5){\circle*{0.1}}
\put(0.5,0.5){\circle*{0.1}} \end{picture}%
\begin{picture}(1,1)(0,0.4) \put(0.5,0.5){\circle{1}} \multiput(0.5,0)(0,0.083){13}{\circle*{0.05}} 
\put(0.15,0.15){\vector(-1,1){0}}
\put(0.15,0.85){\vector(1,1){0}}
\put(0.99,0.5){\vector(0,-1){0}}
\end{picture}
\]
Here double solid line denotes the inverse density propagator ${\cal D}%
\left( {\bf x},{\bf x}^{\prime }\right) $. Therefore the final expression
for the second order correction to $E_{xc}$ is given by, 
\begin{eqnarray*}
E_{xc,2} &=&\setlength{\unitlength}{1cm}\frac{1}{4\beta }\;%
\begin{picture}(1,1)(0,0.4) \put(0.5,0.5){\circle{1}}
\multiput(0.15,0.15)(0.05,0.05){15}{\circle*{0.05}}
\multiput(0.85,0.15)(-0.05,0.05){15}{\circle*{0.05}}
\put(0.01,0.5){\vector(0,-1){0}} \put(0.99,0.5){\vector(0,1){0}}
\put(0.5,1){\vector(-1,0){0}} \put(0.5,0){\vector(1,0){0}} \end{picture}-%
\frac{1}{4\beta }\;\begin{picture}(1,1)(0,0.4) \put(0.5,0.5){\circle{1}}
\multiput(0.85,0.85)(0.08,0){11}{\circle*{0.05}}
\multiput(0.85,0.15)(0.08,0){11}{\circle*{0.05}}
\put(0.01,0.5){\vector(0,-1){0}} \put(0.99,0.5){\vector(0,1){0}}\end{picture}%
\begin{picture}(0.5,1)(0,0.4) \end{picture}{}\begin{picture}(1,1)(0,0.4)
\put(0.5,0.5){\circle{1}} \put(0.01,0.5){\vector(0,-1){0}}
\put(0.99,0.5){\vector(0,1){0}}\end{picture} \\
&&+\setlength{\unitlength}{1cm}\frac{1}{2\beta }\;%
\begin{picture}(1,1)(0,0.4) \put(0.5,0.5){\circle{1}}
\multiput(0.16,0.15)(0,0.07){11}{\circle*{0.05}}
\multiput(0.84,0.15)(0,0.07){11}{\circle*{0.05}}
\put(0.01,0.5){\vector(0,-1){0}}
\put(0.99,0.5){\vector(0,1){0}}\put(0.5,1){\vector(-1,0){0}}
\put(0.5,0){\vector(1,0){0}}\end{picture}+\frac{1}{2\beta }\;%
\begin{picture}(1,1)(0,0.4) \put(0.5,0.5){\circle{1}} \multiput(0.5,0)(0,0.083){13}{\circle*{0.05}} 
\put(0.01,0.5){\vector(0,-1){0}}
\put(0.85,0.15){\vector(1,1){0}}
\put(0.85,0.85){\vector(-1,1){0}}
\end{picture} 
\begin{picture}(0.5,1)(0,0.4) \put(0,0.52){\line(1,0){0.5}}
\put(0,0.47){\line(1,0){0.5}} \put(0,0.5){\circle*{0.1}}
\put(0.5,0.5){\circle*{0.1}} \end{picture}%
\begin{picture}(1,1)(0,0.4) \put(0.5,0.5){\circle{1}} \multiput(0.5,0)(0,0.083){13}{\circle*{0.05}} 
\put(0.15,0.15){\vector(-1,1){0}}
\put(0.15,0.85){\vector(1,1){0}}
\put(0.99,0.5){\vector(0,-1){0}}
\end{picture}
\end{eqnarray*}
It is instructive to apply the above procedure to the case of a homogeneous
electron gas in the zero temperature limit. It can be demonstrated that in
this case Rule \#3 leads to the complete elimination of 2PR graphs. Indeed,
at this limit one can show that 
\[
\setlength{\unitlength}{1cm}%
\begin{picture}(1.5,1)(0,0.4)
\put(0.5,0.5){\circle{1}}
\put(0,0.5){\circle*{1}}
\put(1,0.5){\circle*{0.3}}
\put(0.5,1){\vector(-1,0){0}}
\put(0.5,0){\vector(1,0){0}}
\end{picture}+%
\begin{picture}(1.5,1)(0,0.4) \put(1,0.5){\circle{1}}
\put(1,1){\vector(-1,0){0}}
\put(1,0){\vector(1,0){0}}
\put(0.5,0.5){\circle*{1}}
\end{picture} 
\begin{picture}(0.5,1)(0,0.4) \put(0,0.52){\line(1,0){0.5}}
\put(0,0.47){\line(1,0){0.5}} \put(0,0.5){\circle*{0.1}}
\put(0.5,0.5){\circle*{0.1}} \end{picture}%
\begin{picture}(1.5,1)(0,0.4)
\put(0.5,0.5){\circle{1}} 
\put(1,0.5){\circle*{0.3}}
\put(0.5,1){\vector(-1,0){0}}
\put(0.5,0){\vector(1,0){0}}
\end{picture}=0.
\]
Here, black circles denote parts of the diagram that are connected to each
other via two propagators. For example, in the expression for the second
order correction to $E_{xc}$ the last two graphs completely cancel each
other $\ $%
\[
\setlength{\unitlength}{1cm}%
\begin{picture}(1,1)(0,0.4) \put(0.5,0.5){\circle{1}} \multiput(0.15,0.15)(0,0.05){15}{\circle*{0.05}} \multiput(0.85,0.15)(0,0.05){15}{\circle*{0.05}} \put(0.01,0.5){\vector(0,-1){0}}
\put(0.99,0.5){\vector(0,1){0}}\put(0.5,1){\vector(-1,0){0}}
\put(0.5,0){\vector(1,0){0}}\end{picture}+%
\begin{picture}(1,1)(0,0.4) \put(0.5,0.5){\circle{1}} \multiput(0.5,0)(0,0.05){21}{\circle*{0.05}} 
\put(0.01,0.5){\vector(0,-1){0}}
\put(0.85,0.15){\vector(1,1){0}}
\put(0.85,0.85){\vector(-1,1){0}}
\end{picture} 
\begin{picture}(0.5,1)(0,0.4) \put(0,0.52){\line(1,0){0.5}}
\put(0,0.47){\line(1,0){0.5}} \put(0,0.5){\circle*{0.1}}
\put(0.5,0.5){\circle*{0.1}} \end{picture}%
\begin{picture}(1,1)(0,0.4) \put(0.5,0.5){\circle{1}} \multiput(0.5,0)(0,0.05){21}{\circle*{0.05}} 
\put(0.15,0.15){\vector(-1,1){0}}
\put(0.15,0.85){\vector(1,1){0}}
\put(0.99,0.5){\vector(0,-1){0}}
\end{picture}=0,
\]
and as expected,\cite{DFTbook1} 
\[
E_{xc,2}^{\hom }\ =\setlength{\unitlength}{1cm}\frac{1}{4\beta }\;%
\begin{picture}(1,1)(0,0.4) \put(0.5,0.5){\circle{1}}
\multiput(0.15,0.15)(0.05,0.05){15}{\circle*{0.05}}
\multiput(0.85,0.15)(-0.05,0.05){15}{\circle*{0.05}}
\put(0.01,0.5){\vector(0,-1){0}} \put(0.99,0.5){\vector(0,1){0}}
\put(0.5,1){\vector(-1,0){0}} \put(0.5,0){\vector(1,0){0}} \end{picture}-%
\frac{1}{4\beta }\;\begin{picture}(1,1)(0,0.4) \put(0.5,0.5){\circle{1}}
\multiput(0.85,0.85)(0.08,0){11}{\circle*{0.05}}
\multiput(0.85,0.15)(0.08,0){11}{\circle*{0.05}}
\put(0.01,0.5){\vector(0,-1){0}} \put(0.99,0.5){\vector(0,1){0}}\end{picture}%
\begin{picture}(0.5,1)(0,0.4) \end{picture}{}\begin{picture}(1,1)(0,0.4)
\put(0.5,0.5){\circle{1}} \put(0.01,0.5){\vector(0,-1){0}}
\put(0.99,0.5){\vector(0,1){0}}\end{picture}.
\]
Local Density Approximation (LDA) represents a popular choice for $%
E_{xc}\left[ n\right] $ in first-principles calculations. This approximation
and subsequent corrections can obtained via the derivative expansion\cite
{ryder} of the exchange-correlation functional 
\[
E_{xc}\left[ n\right] =\int \left( E_{xc}^{\left( 0\right) }\left( n\left( 
{\bf x}\right) \right) +E_{xc}^{\left( 2\right) }\left( n\left( {\bf x}%
\right) \right) \left( \nabla n\left( {\bf x}\right) \right) ^2+\ldots
\right) d{\bf x}
\]
where $E_{xc}^{\left( k\right) }$ is a {\it function} of $n\left( {\bf x}%
\right) ,$ not a functional{\bf .}

LDA corresponds to the first term in this expansion; 
\[
E_{xc}^{LDA}\left[ n\right] \equiv \int E_{xc}^{\left( 0\right) }\left(
n\left( {\bf x}\right) \right) d{\bf x}. 
\]
Function $E_{xc}^{\left( 0\right) }\left( n\left( {\bf x}\right) \right) $
can be found by evaluating the above expansion at constant density $n\left( 
{\bf x}\right) =n_0$ 
\[
E_{xc}^{\left( 0\right) }\left( n_0\right) =\frac 1VE_{xc}\left[ n_0\right] ,%
\hspace{0.5cm}n_0=\text{const.} 
\]
Here $E_{xc}\left[ n_0\right] $ represents the exchange-correlation energy
of the homogeneous electron gas with density $n_0$.

Regarding the results obtained in this section, we would like to emphasize the
following points. Not only does the effective action formalism leads to a
straightforward set of rules to calculate the exchange-correlation
functional up to any arbitrary order, but it can also be used to generate
similar quantities for descriptions based on the observables other than
particle density (for example, current-density functional theory). One
should also remember that the expansion of $E_{xc}\left[ n\right] $
represents only part of the general picture provided by the effective action
formalism. As we show later, the same formalism allows us to develop a
rigorous and systematic Kohn-Sham theory for one-electron propagators and
many-body excitation energies.

\subsection{Kohn-Sham self-consistent procedure}

As we have demonstrated earlier, application of the variational principle
leads to the Kohn-Sham self-consistent procedure. In the case of the
traditional density-functional theory, the typical self-consistent procedure
takes the form

\begin{enumerate}
\item  Start with some reasonable guess for the Kohn-Sham potential $%
J_0\left( {\bf x}\right) $.

\item  Solve Kohn-Sham single-particle equations.

\item  Determine new Kohn-Sham potential $J_0\left( {\bf x}\right) $ using 
\[
J_0\left( {\bf x}\right) =\int \frac{n\left( {\bf x}^{\prime }\right) }{%
\left| {\bf x}-{\bf x}^{\prime }\right| }d{\bf x}^{\prime }+\upsilon
_{xc}\left( {\bf x}\right) , 
\]
where the exchange-correlation potential $\upsilon _{xc}\left( {\bf x}%
\right) $ is defined as, 
\[
\upsilon _{xc}\left( {\bf x}\right) =\frac{\delta E_{xc}\left[ n\right] }{%
\delta n\left( {\bf x}\right) }. 
\]

\item  Repeat from step 2 until self-consistency is achieved.
\end{enumerate}

In LDA the exchange-correlation functional is an explicit functional of
electron density $n\left( {\bf x}\right) ,$ and the exchange-correlation
potential $\upsilon _{xc}\left( {\bf x}\right) $ can be obtained by
straightforward differentiation. In the case of the diagrammatic expansion,
this simple property no longer holds and $E_{xc}\left[ n\right] $ appears as
an implicit functional of $n\left( {\bf x}\right) $. The
exchange-correlation potential $\upsilon _{xc}\left( {\bf x}\right) $ can
still be found using the equation: 
\[
\upsilon _{xc}\left( {\bf x}\right) =-\int {\cal D}\left( {\bf x},{\bf x}%
^{\prime }\right) \frac{\delta E_{xc}\left[ n\right] }{\delta J_0\left( {\bf %
x}^{\prime }\right) }d{\bf x}^{\prime },
\]
The functional derivative in the above expression can be easily evaluated
based on the following relationships: 
\[
\frac{\delta {\cal G}_0\left( x_1,x_2\right) }{\delta J_0\left( {\bf x}%
\right) }=\int_0^\beta {\cal G}_0\left( x_1,x\right) {\cal G}_0\left(
x,x_2\right) d\tau ,
\]
\[
\frac{\delta {\cal D}\left( {\bf x},{\bf x}^{\prime }\right) }{\delta {\cal G%
}_0\left( x_1,x_2\right) }=2{\cal D}\left( {\bf x},{\bf x}_1\right) {\cal G}%
_0\left( x_1,x_2\right) {\cal D}\left( {\bf x}_2,{\bf x}^{\prime }\right) .
\]
For example, the first order correction to $\upsilon _{xc}\left( {\bf x}%
\right) $ is given by, 
\[
\upsilon _{xc,1}\left( {\bf x}\right) =\frac{1}{\beta }\;\setlength{%
\unitlength}{1cm}%
\begin{picture}(0.5,1)(0,0.4) \put(0,0.52){\line(1,0){0.5}}
\put(0,0.47){\line(1,0){0.5}} \put(0,0.5){\circle*{0.1}}
\put(0.5,0.5){\circle*{0.1}} \end{picture}%
\begin{picture}(1,1)(0,0.4)
\put(0.5,0.5){\circle{1}} \multiput(0.5,0)(0,0.083){13}{\circle*{0.05}}
\put(0.15,0.15){\vector(-1,1){0}}
\put(0.15,0.85){\vector(1,1){0}}
\put(0.99,0.5){\vector(0,-1){0}}\end{picture}
\]
Obviously, even in case of the diagrammatic expansion of $E_{xc}$ one could
still use the above mentioned self-consistent procedure. To reduce the
computational effort, however, slight modification of that procedure might
be advantageous. Namely, we suggest to shift the emphasis from density $n$
to Kohn-Sham potential $J_0$. Indeed, Corollary 1 guarantees that there is a
one-to-one correspondence between $n$ and $J_0.$ Thus, we can consider the
effective action functional that depends on $J_0$ rather than $n$: 
\[
\overline{\Gamma }\left[ J_0\right] \equiv \Gamma \left[ n\left[ J_0\right]
\right] .
\]
The variational principle then takes the form 
\[
\frac{\delta \overline{\Gamma }\left[ J_0\right] }{\delta J_0\left( {\bf x}%
\right) }=0.
\]
In other words, one has to find a Kohn-Sham potential that minimizes the
effective action functional $\overline{\Gamma }\left[ J_0\right] $. To
accomplish this task, one could use the so-called steepest descent
minimization method.\cite{rao} In this case the self-consistent procedure
takes the form:

\begin{enumerate}
\item  Start with some reasonable guess for the Kohn-Sham potential $%
J_0\left( {\bf x}\right) $.

\item  Calculate the direction of steepest descent $s\left( {\bf x}\right) $
as, 
\[
s\left( {\bf x}\right) =-\frac{\delta \overline{\Gamma }\left[ J_0\right] }{%
\delta J_0\left( {\bf x}\right) }
\]
or 
\[
s\left( {\bf x}\right) =-\int {\cal D}^{-1}\left( {\bf x,x}^{\prime }\right)
J_0\left( {\bf x}^{\prime }\right) d{\bf x}^{\prime }-\frac{\delta \overline{%
\Gamma }_{int}\left[ J_0\right] }{\delta J_0\left( {\bf x}\right) }
\]

\item  Determine new Kohn-Sham potential $J_0^{new}\left( {\bf x}\right) $
from the old Kohn-Sham potential $J_0^{old}\left( {\bf x}\right) $ by
stepping along the direction of steepest descent: 
\[
J_0^{new}\left( {\bf x}\right) =J_0^{old}\left( {\bf x}\right) +\alpha
s\left( {\bf x}\right) 
\]
where $\alpha $ is the length of the step.

\item  Repeat from step 2 until self-consistency is achieved.
\end{enumerate}

The advantage of this self-consistent procedure is that it is much easier to
calculate 
\[
\frac{\delta \overline{\Gamma }\left[ J_0\right] }{\delta J_0\left( {\bf x}%
\right) }
\]
rather than 
\[
\frac{\delta \Gamma \left[ n\right] }{\delta n\left( {\bf x}\right) }.
\]
For example, when $\overline{\Gamma }_{int}\left[ J_0\right] $ is
approximated by its first order correction 
\[
\overline{\Gamma }_{int}\left[ J_{0}\right] \approx \setlength{%
\unitlength}{1cm}\frac{1}{2\beta }{}\;%
\begin{picture}(1,1)(0,0.4) \put(0.5,0.5){\circle{1}}
\put(0.01,0.5){\vector(0,-1){0}} \end{picture}{}%
\begin{picture}(0.5,1)(0,0.4) 
\multiput(0,0.5)(0.082,0){7}{\circle*{0.05}}
\end{picture}{}%
\begin{picture}(1,1)(0,0.4) \put(0.5,0.5){\circle{1}}
\put(0.99,0.5){\vector(0,-1){0.0}} \end{picture}-\frac{1}{2\beta }{}\;%
\begin{picture}(1,1)(0,0.4) \put(0.5,0.5){\circle{1}}
\put(0.01,0.5){\vector(0,-1){0}} \put(0.99,0.5){\vector(0,1){0}}
\multiput(0.5,0)(0,0.083){13}{\circle*{0.05}}
\end{picture}
\]
the steepest descent direction is given by 
\begin{eqnarray*}
s &&\left( {\bf x}\right) =-\int {\cal D}^{-1}\left( {\bf x,x}^{\prime
}\right) \left( J_0\left( {\bf x}^{\prime }\right) -\int \frac{n\left( {\bf x%
}^{\prime \prime }\right) }{\left| {\bf x}^{\prime \prime }-{\bf x}^{\prime
}\right| }d{\bf x}^{\prime \prime }\right) d{\bf x}^{\prime } \\
&&+\int_0^\beta d\tau \int \int {\cal G}_0({\bf y}0,{\bf x}\tau ){\cal G}_0(%
{\bf x}\tau ,{\bf y}^{\prime }0)\frac{n({\bf y}^{\prime },{\bf y})}{\left| 
{\bf y}^{\prime }-{\bf y}\right| }d{\bf y}d{\bf y}^{\prime }.
\end{eqnarray*}
The advantage of this expression as compared to the first order correction
to $\upsilon _{xc,1}$ is that here, we avoid the calculation of the inverse
density propagator ${\cal D}\left( {\bf x},{\bf x}^{\prime }\right) $. Note
that ${\cal D}^{-1}\left( {\bf x,x}^{\prime }\right) $ can easily be written
in terms of Kohn-Sham single-particle orbitals and energies (see Eqn $\left( 
\ref{D(x,x')}\right) $)$.$

\section{Time-dependent probe}

To study excitation energies and one-electron propagators, it is necessary
to consider an imaginary-time-dependent probe. The definition of the
functional $W\left[ J\right] $ is changed correspondingly 
\begin{equation}
e^{-W\left[ J\right] }=\int D\psi ^{\dagger }D\psi \;e^{-S\left[ \psi
^{\dagger },\psi \right] -\int J\left( x\right) \psi ^{\dagger }\left(
x\right) \psi \left( x\right) dx}.  \label{tdgenfunc}
\end{equation}
Note that the parameter $\beta $ has been absorbed into $W\left[ J\right] $.
In order to proceed with the inversion method, we need to assure that the
map $J\left( x\right) \rightarrow n\left( x\right) ,$%
\begin{equation}
n\left( x\right) =\frac{\delta W\left[ J\right] }{\delta J\left( x\right) },
\label{tdJ}
\end{equation}
is invertible. Since at the end the time-dependent probe is set to zero, we
can assume our source to be infinitesimally small. Therefore it suffices to
prove the invertibilty in the small neighborhood of {\it time-independent}
external source: 
\[
J\left( {\bf x}\right) +\delta J\left( x\right) \rightarrow n\left( {\bf x}%
\right) +\delta n\left( x\right) . 
\]
Since, 
\[
\delta n\left( x\right) =\int \left( \frac{\delta ^2W\left[ J\right] }{%
\delta J\left( x\right) \delta J\left( x^{\prime }\right) }\right) _{J\left( 
{\bf x}\right) }\delta J\left( x^{\prime }\right) dx^{\prime }, 
\]
we need to show that the operator 
\[
W^{\left( 2\right) }\left( x,x^{\prime }\right) \equiv \left( \frac{\delta
^2W\left[ J\right] }{\delta J\left( x\right) \delta J\left( x^{\prime
}\right) }\right) _{J\left( {\bf x}\right) } 
\]
has no zero eigenvalues. This property follows from the following theorem.

{\bf Theorem 2: }{\it The operator }$W^{\left( 2\right) }\left( x,x^{\prime
}\right) ${\it \ is strictly negative definite. }(The proof is given in the
appendix.)

This theorem guarantees that as long as two infinitesimally small probes
differ by more than a pure time-dependent function, they would produce two
different densities. Once this one-to-one correspondence has been
established, further analysis proceeds similar to the time-independent case.
The effective action functional is defined as, 
\[
\Gamma \left[ n\right] =W\left[ J\right] -\int J\left( x\right) n\left(
x\right) dx, 
\]
where $J$ is assumed to be a functional of $n$ by Eqn $\left( \ref{tdJ}%
\right) $. Using the inversion method,\cite{okumura} the effective action
functional can be found as a power series in terms of the coupling constant: 
\[
\Gamma \left[ n,e^{2}\right] =\Gamma _{0}\left[ n\right] +e^{2}\Gamma
_{1}\left[ n\right] +e^{4}\Gamma _{2}\left[ n\right] +\ldots . 
\]
Zeroth order term is given by,

\[
\Gamma _0\left[ n\right] =W_0\left[ J_0\right] -\int J_0\left( x\right)
n\left( x\right) dx. 
\]
The functional $W_0\left[ J_0\right] $ describes a Kohn-Sham system of
noninteracting electrons in the presence of an imaginary-time-dependent
external potential $J_0\left( x\right) .$ Kohn-Sham potential $J_0\left(
x\right) $ is chosen such that the time-dependent density $n\left( x\right) $
is reproduced. Diagrammatic structure of $\Gamma _i\left[ n\right] $ is the
same as its time-independent counterparts with the only difference being
that Kohn-Sham propagator is now defined in the presence of a time-dependent
Kohn-Sham potential $J_0\left( x\right) ,$%
\[
{\cal G}_0^{-1}\left( x,x^{\prime }\right) =-\left( \frac \partial {\partial
\tau }-\frac{\nabla ^2}{2m}+\upsilon _{ion}({\bf x})+J_0\left( x\right)
\right) \delta (x-x^{\prime }), 
\]
and the inverse density propagator becomes 
\[
{\cal D}\left( x,x^{\prime }\right) =-\left[ {\cal G}_0\left( x,x^{\prime
}\right) {\cal G}_0\left( x^{\prime },x\right) \right] ^{-1}. 
\]

\section{One-electron propagators}

Using the effective action formalism it is possible to develop a systematic
Kohn-Sham density-functional approach to one-electron propagators. It is
common practice to use converged Kohn-Sham single-particle orbitals and
energies in quasiparticle calculations. However, very often these methods
are not very systematic in their use of Kohn-Sham based quantities.\cite{GW1}
The formalism presented below provides a rigorous theoretical foundation for
the calculation of quasiparticle properties based on Kohn-Sham
noninteracting system.

Consider $W\left[ J\right] $ in the presence of the auxiliary nonlocal
source $\xi \left( x,x^{\prime }\right) $ 
\[
e^{-W\left[ J\right] }=\int D\psi ^{\dagger }D\psi \;e^{-S_{\xi }\left[ \psi
^{\dagger },\psi \right] }, 
\]
where 
\begin{eqnarray*}
S_{\xi }\left[ \psi ^{\dagger },\psi \right] &=&S\left[ \psi ^{\dagger
},\psi \right] +\int J\left( x\right) \psi ^{\dagger }\left( x\right) \psi
\left( x\right) dx \\
&&+\int \int \xi \left( x,x^{\prime }\right) \psi ^{\dagger }\left( x\right)
\psi \left( x^{\prime }\right) dxdx^{\prime }.
\end{eqnarray*}
The nonlocal source, $\xi \left( x,x^{\prime }\right) ,$ allows us to write
the one-electron propagator or the (finite temperature) Green's function, $%
{\cal G}(x,x^{\prime })=-\left\langle T_{\tau }\psi \left( x\right) \psi
^{\dagger }\left( x^{\prime }\right) \right\rangle $, as a functional
derivative

\[
{\cal G}(x,x^{\prime })=\left( \frac{\delta W\left[ J\right] }{\delta \xi
\left( x^{\prime },x\right) }\right) _{J}. 
\]
Using the well-known property of the Legendre transformation,\cite{justin}

\begin{equation}
\left( \frac{\delta W\left[ J\right] }{\delta \xi \left( x^{\prime
},x\right) }\right) _J=\left( \frac{\delta \Gamma \left[ n\right] }{\delta
\xi \left( x^{\prime },x\right) }\right) _n,  \label{property}
\end{equation}
the one-electron propagator can be expressed in terms of the effective
action functional as,

\begin{equation}
{\cal G}(x,x^{\prime })=\left( \frac{\delta \Gamma _{0}\left[ n\right] }{%
\delta \xi \left( x^{\prime },x\right) }\right) _{n}+\left( \frac{\delta
\Gamma _{int}\left[ n\right] }{\delta \xi \left( x^{\prime },x\right) }%
\right) _{n}.  \label{Green}
\end{equation}
Using the property (\ref{property}) we obtain that 
\begin{equation}
\left( \frac{\delta \Gamma _{0}\left[ n\right] }{\delta \xi \left( x^{\prime
},x\right) }\right) _{n}=\left( \frac{\delta W_{0}\left[ J_{0}\right] }{%
\delta \xi \left( x^{\prime },x\right) }\right) _{J_{0}}={\cal G}%
_{0}(x,x^{\prime }).  \label{firstterm}
\end{equation}
Let us consider the second term in Eqn $\left( \ref{Green}\right) $. One can
show that 
\begin{eqnarray}
\left( \frac{\delta \Gamma _{int}\left[ n\right] }{\delta \xi \left(
x^{\prime },x\right) }\right) _{n} &=&\int \int \frac{\delta \Gamma
_{int}\left[ n\right] }{\delta {\cal G}_{0}\left( y,y^{\prime }\right) }%
{\cal G}_{0}(y,x^{\prime }){\cal G}_{0}(x,y^{\prime })dydy^{\prime } 
\nonumber \\
&&\ -\int \frac{\delta \Gamma _{int}\left[ n\right] }{\delta n(y)}{\cal G}%
_{0}(y,x^{\prime }){\cal G}_{0}(x,y)dy.  \label{second}
\end{eqnarray}
Our original system is recovered by setting $\xi $ to zero. Using Eqns (\ref
{Green}), (\ref{firstterm}), (\ref{second}) the exact one-electron
propagator (in operator notation) is given by,

\begin{equation}
{\cal G}={\cal G}_0+{\cal G}_0\cdot \Sigma _0\cdot {\cal G}_0,  \label{Dyson}
\end{equation}
where the Kohn-Sham self-energy $\Sigma _0$ is given by,

\begin{equation}
\Sigma _0\left( x_1,x_2\right) =\frac{\delta \Gamma _{int}\left[ n\right] }{%
\delta {\cal G}_0\left( x_2,x_1\right) }-J_0(x_1)\delta (x_1-x_2)
\label{ksself}
\end{equation}
and $J_0(x_1)$ is the Kohn-Sham potential,

\[
J_0(x)=\frac{\delta \Gamma _{int}\left[ n\right] }{\delta n(x)}. 
\]
The functional derivative in Eqn (\ref{ksself}) can be easily evaluated
since the functional $\Gamma _{int}$ can be expressed entirely in terms of
Kohn-Sham Green's functions ${\cal G}_0$. The above expression for the
one-electron propagator can also be written as,

\[
{\cal G}={\cal G}_0+{\cal G}_0\cdot \widetilde{\Sigma }_0\cdot {\cal G}, 
\]
where

\[
\widetilde{\Sigma }_{0}=\Sigma _{0}\cdot (1+{\cal G}_{0}\cdot \Sigma
_{0})^{-1}=\left( \Sigma _{0}^{-1}+{\cal G}_{0}\right) ^{-1}. 
\]
Therefore the exact self-energy $\Sigma $ is given by,

\[
\Sigma \left( x,x^{\prime }\right) =J_{0}(x)\delta (x-x^{\prime })+%
\widetilde{\Sigma }_{0}(x,x^{\prime }). 
\]

The above formulation provides a systematic way to study one-electron
propagators and self-energy in Kohn-Sham density-functional theory. The
important feature of this formulation is that both self-energy and Kohn-Sham
potential are determined from one quantity $\Gamma _{int}\left[ n\right] $. 
{\it A single approximation to the functional }$\Gamma _{int}\left[ n\right] 
${\it \ simultaneously generates both the self-energy and Kohn-Sham
potential.} This is to be contrasted with the common strategy of performing 
{\it separate} approximations for Kohn-Sham potential and the self-energy.%
\cite{GW1}

\section{Excitation energies}

In addition to one-electron propagators, the effective action formalism
allows a systematic study of the many-body excited states in
density-functional theory.\cite{fukuda} Consider the Fourier transform of $%
W^{\left( 2\right) }\left( x,x^{\prime }\right) :$%
\begin{eqnarray*}
W^{\left( 2\right) }\left( {\bf x},{\bf x}^{\prime },i\nu _s\right)
&=&\int_0^\beta W^{\left( 2\right) }\left( x,x^{\prime }\right) e^{i\nu
_s\left( \tau -\tau ^{\prime }\right) }d\left( \tau -\tau ^{\prime }\right) ,
\\
\nu _s &=&2\pi s/\beta ,
\end{eqnarray*}
It can be analytically continued into the complex plane $\omega $: 
\[
W^{\left( 2\right) }\left( {\bf x},{\bf x}^{\prime },\omega \right)
=W^{\left( 2\right) }\left( {\bf x},{\bf x}^{\prime },i\nu _s\right) |_{i\nu
_s\rightarrow \omega +i\eta }. 
\]
The proposition below guarantees that this analytic continuation has an
inverse in the upper complex plane $\omega $ including the real axis.

{\bf Proposition 2}: {\it The operator }$W^{\left( 2\right) }{\bf (x},{\bf x}%
^{\prime },\omega )${\it \ has no zero eigenvalues when }$\omega ${\it \ is
located in the upper half of the complex plane\ including the real axis. }%
(see appendix for the proof.)

Let us define the excitation kernel as, 
\[
\Gamma ^{\left( 2\right) }(x,x^{\prime })=\left( \frac{\delta \Gamma
^2\left[ n\right] }{\delta n\left( x\right) \delta n\left( x^{\prime
}\right) }\right) _{n\left( {\bf x}\right) }. 
\]
It easy to show that 
\[
\int W^{\left( 2\right) }(x,x^{\prime })\Gamma ^{\left( 2\right) }(x^{\prime
},y)dx^{\prime }=-\delta \left( x-y\right) , 
\]
or in terms of Fourier transforms 
\begin{equation}
\int W^{\left( 2\right) }{\bf (x},{\bf x}^{\prime },i\nu _s)\Gamma ^{\left(
2\right) }({\bf x}^{\prime },{\bf y,-}i\nu _s)d{\bf x}^{\prime }=-\delta
\left( {\bf x}-{\bf y}\right) .  \label{WG=1}
\end{equation}
{\bf Proposition 3:} $\Gamma ^{\left( 2\right) }({\bf x},{\bf y,-}i\nu _s)$
has a unique analytic continuation $\Gamma ^{\left( 2\right) }({\bf x},{\bf %
y,-}\omega )$ such that

\begin{enumerate}
\item  it does not have any zeros in the upper half of the complex plane $%
\omega $ including the real axis.

\item  $\left[ \Gamma ^{\left( 2\right) }({\bf x},{\bf y,-}\omega )\right]
^{-1}\sim \frac{1}{\left| \omega \right| },$ $\left( \omega \rightarrow
\infty \right) $.
\end{enumerate}

The proof is given in the appendix. Based on the above statement we can
extend the relationship $\left( \ref{WG=1}\right) $ to the whole complex
plane $\omega :$%
\begin{equation}
\int W^{\left( 2\right) }{\bf (x},{\bf x}^{\prime },\omega )\Gamma ^{\left(
2\right) }({\bf x}^{\prime },{\bf y,-}\omega )d{\bf x}^{\prime }=-\delta
\left( {\bf x}-{\bf y}\right) .  \label{WGw=1}
\end{equation}
Consider $W^{\left( 2\right) }{\bf (x},{\bf x}^{\prime },\omega )$ in the
zero temperature limit $\left( \beta \rightarrow \infty \right) $. It is
well-known that it has poles just below the real axis at the exact
excitation energies, 
\[
\omega =E_l-E_0-i\eta . 
\]
Then it follows from Eqn $\left( \ref{WGw=1}\right)$ that $\Gamma ^{\left(
2\right) }({\bf x},{\bf y,-}\omega)$ has zero eigenvalues at the exact
excitation energies.\cite{fukuda} In other words, when $\omega =E_l-E_0-i\eta
$ , there exists $\xi \left( {\bf x}\right)$ such that,\cite{fukuda} 
\begin{equation}
\int \Gamma ^{\left( 2\right) }({\bf x},{\bf y,-}\omega )\xi \left( {\bf x}%
\right) d{\bf x}=0.  \label{excitation}
\end{equation}
Using the coupling constant expansion for $\Gamma (n)$ the excitation kernel
becomes

\[
\Gamma ^{\left( 2\right) }({\bf x},{\bf y,-}i\nu _{s})=\Gamma _{0}^{\left(
2\right) }({\bf x},{\bf y,-}i\nu _{s})+\Gamma _{int}^{\left( 2\right) }({\bf %
x},{\bf y,-}i\nu _{s}). 
\]
Substituting this in Eqn $\left( \ref{excitation}\right) $ and using the
fact that, 
\[
\left[ \Gamma _{0}^{\left( 2\right) }({\bf x},{\bf x}^{\prime },-\omega
)\right] ^{-1}=-W_{0}^{\left( 2\right) }({\bf x},{\bf x}^{\prime },\,\omega
), 
\]
we obtain 
\begin{equation}
\xi \left( {\bf x}\right) =\int \int W_{0}^{\left( 2\right) }({\bf x},{\bf x}%
^{\prime },\,\omega )\Gamma _{int}^{\left( 2\right) }({\bf x}^{\prime },{\bf %
y}^{\prime }{\bf ,-}\omega )\xi \left( {\bf y}^{\prime }\right) d{\bf x}%
^{\prime }d{\bf y}^{\prime }.  \label{excit-energy2}
\end{equation}
Here $W_{0}^{\left( 2\right) }({\bf x},{\bf x}^{\prime },\omega )$
represents the negative of the density-correlation function for Kohn-Sham
noninteracting system;

\[
W_0^{\left( 2\right) }{\bf (x},{\bf x}^{\prime },\omega )=\sum_l\frac{%
n_l^{*}\left( {\bf x}\right) n_l\left( {\bf x}^{\prime }\right) }{\omega
-\omega _l^{ks}+i\eta }-\frac{n_l\left( {\bf x}\right) n_l^{*}\left( {\bf x}%
^{\prime }\right) }{\omega +\omega _l^{ks}+i\eta }, 
\]
where 
\[
\left\langle l^{ks}\left| \ \widehat{n}\left( {\bf x}\right) \right|
0^{ks}\right\rangle =n_l\left( {\bf x}\right) , 
\]
and 
\[
\omega _l^{ks}=E_l^{ks}-E_0^{ks}. 
\]
Here all the quantities refer to the Kohn-Sham noninteracting system. Aside
from the notational differences, Eqn $\left( \ref{excit-energy2}\right) $
coincides with similar expressions derived using time-dependent DFT.\cite
{peter,excitation1} Searching for a solution of the form 
\[
\xi \left( {\bf x}\right) =\sum a_l n_l\left( {\bf x}\right)
+\sum b_ln_l^{*}\left( {\bf x}\right) , 
\]
we obtain the following matrix equation,\cite{excitation1} 
\[
\left[ 
\begin{array}{cc}
L & M \\ 
M^{*} & L^{*}
\end{array}
\right] \left[ 
\begin{array}{c}
A \\ 
B
\end{array}
\right] =\left( \omega +i\eta \right) \left[ 
\begin{array}{cc}
-1 & 0 \\ 
0 & 1
\end{array}
\right] \left[ 
\begin{array}{c}
A \\ 
B
\end{array}
\right] , 
\]
where 
\[
M_{ij}=\int \int n_i^{*}\left( {\bf x}\right) \Gamma
_{int}^{\left( 2\right) }({\bf x},{\bf y,-}\omega )n_j^{*}\left( 
{\bf y}\right) d{\bf x}d{\bf y,} 
\]
\begin{eqnarray*}
L_{ij} &=&\int \int n_i^{*}\left( {\bf x}\right) \Gamma
_{int}^{\left( 2\right) }({\bf x},{\bf y,-}\omega )n_j\left( 
{\bf y}\right) d{\bf x}d{\bf y} \\
&&\ {\bf +}\left( E_i-E_0\right) \delta _{ij},
\end{eqnarray*}
and 
\[
\left( A\right) _i=a_i,\;\left( B\right) _i=b_i. 
\]
The values of $\omega $ for which the above matrix equation has nontrivial
solutions, is determined by the condition 
\[
\det \left( \left[ 
\begin{array}{cc}
L & M \\ 
M^{*} & L^{*}
\end{array}
\right] -\left( \omega +i\eta \right) \left[ 
\begin{array}{cc}
-1 & 0 \\ 
0 & 1
\end{array}
\right] \right) =0. 
\]
Therefore, the effective action formalism presents an alternative way (as
compared to time-dependent density-functional theory) for calculating the
excitation energies. However, in addition to it, the effective action
formalism also provides a means of calculating the exchange-correlation
kernel. This feature is missing in conventional time-dependent
density-functional theory.\cite{GH}

\section{Concluding Remarks.}

Based on the effective action formalism\cite{fukuda} we developed a rigorous
formulation of generalized Kohn-Sham theory. This formulation is
specifically geared towards practical calculations of the ground and excited
properties of real systems. Indeed, in the case of a particle-density based
description of nonrelativistic many-electron system we arrived at a
systematic way to study the exchange-correlation functional, one-electron
propagators and many-body excitation energies entirely in terms Kohn-Sham
single-particle orbitals and energies. The presented formalism is very
general and can be applied to a various many-body systems for constructing
Kohn-Sham like description in terms of the expectation value of any general
operator (e.g. spin-density, current-density-functional theory).

\section{Acknowledgments.}

This work was supported in part by the University of Connecticut Research
Foundation.

\appendix 

\section{Properties of functionals $W\left[ J\right] $ and $\Gamma \left[
Q\right] .$}

\subsection{Time-independent probe}

{\bf Theorem 1}: {\it The functional }$W\left[ J\right] =-\frac 1\beta \ln $%
Tr$\left( e^{-\beta \left( \hat H+J\left( 1\right) \hat Q\left( 1\right)
\right) }\right) ${\it \ is strictly concave, i.e. for any }$\alpha ${\it , }%
$0<\alpha <1${\it , and }$J\neq J^{\prime }$

\[
W\left[ \alpha J+\left( 1-\alpha \right) J^{\prime }\right] >\alpha W\left[
J\right] +\left( 1-\alpha \right) W\left[ J^{\prime }\right] 
\]

{\bf Proof:}\cite{israel}

Consider two Hermitian operators $\hat A$ and $\hat B$. Let $\left\{ \Psi
_i\right\} $ be a complete set of eigenstates of the operator $\alpha \hat A%
+\left( 1-\alpha \right) \hat B.$ Then 
\begin{eqnarray*}
\text{Tr}\left( e^{\alpha \hat A+\left( 1-\alpha \right) \hat B}\right)
&=&\sum_i\left\langle \Psi _i\left| e^{\alpha \hat A+\left( 1-\alpha \right) 
\hat B}\right| \Psi _i\right\rangle \\
&=&\sum_ie^{\alpha \left\langle \Psi _i\left| \hat A\right| \Psi
_i\right\rangle +\left( 1-\alpha \right) \left\langle \Psi _i\left| \hat B%
\right| \Psi _i\right\rangle }.
\end{eqnarray*}
Using H\"older's inequality\cite{holder} $\left( a_i,b_i\geq 0\right) ,$ 
\[
\sum_ia_i^\alpha b_i^{1-\alpha }\leq \left( \sum_ia_i\right) ^\alpha \left(
\sum_ib_i\right) ^{1-\alpha }, 
\]
we obtain that 
\[
\text{Tr}\left( e^{\alpha \hat A+\left( 1-\alpha \right) \hat B}\right) \leq
\left( \sum_ie^{\left\langle \Psi _i\left| \hat A\right| \Psi
_i\right\rangle }\right) ^\alpha \left( \sum_ie^{\left\langle \Psi _i\left| 
\hat B\right| \Psi _i\right\rangle }\right) ^{\left( 1-\alpha \right) } 
\]
From H\"older's inequality it follows that the equality in the above
expression holds only if for any $i$%
\begin{equation}
\left\langle \Psi _i\left| \hat A\right| \Psi _i\right\rangle =\left\langle
\Psi _i\left| \hat B\right| \Psi _i\right\rangle +\chi ,  \label{condition1}
\end{equation}
where $\chi $ is constant independent of $i$.

Since $e^x$ is a convex function, 
\begin{eqnarray}
\sum_ie^{\left\langle \Psi _i\left| \hat A\right| \Psi _i\right\rangle }
&\leq &\sum_i\left\langle \Psi _i\left| e^{\hat A}\right| \Psi
_i\right\rangle ,  \nonumber \\
\sum_ie^{\left\langle \Psi _i\left| \hat B\right| \Psi _i\right\rangle }
&\leq &\sum_i\left\langle \Psi _i\left| e^{\hat B}\right| \Psi
_i\right\rangle .  \label{condition2}
\end{eqnarray}
Equal sign in the above equations holds only if $\left\{ \Psi _i\right\} $
are the eigenstates $\hat A$ and $\hat B$ . Collecting all the results
together we obtain the following inequality: 
\begin{equation}
\text{Tr}\left( e^{\alpha \hat A+\left( 1-\alpha \right) \hat B}\right) \leq
\left( \text{Tr}e^{\hat A}\right) ^\alpha \left( \text{Tr}e^{\hat B}\right)
^{\left( 1-\alpha \right) }.  \label{result}
\end{equation}
Equality holds here if and only if operators $\hat A$ and $\hat B$ differ by
a constant: 
\begin{equation}
\hat A=\hat B+const.  \label{condition3}
\end{equation}
Indeed:

$\Rightarrow )$ Suppose Eqn \ref{condition3} is true, then equal sign in $%
\left( \ref{result}\right) $ is obvious.

$\Leftarrow )$ If there is an equal sign in $\left( \ref{result}\right) $
then operators $\hat A$ and $\hat B$ must have a common set of eigenstates $%
\left\{ \Psi _i\right\} $ and condition $\left( \ref{condition1}\right) $
must be true. Then considering the representation of operators $\hat A$ and $%
\hat B$ in the basis of eigenfunctions $\left\{ \Psi _i\right\} $ we obtain 
\[
\hat A=\hat B+const. 
\]
Therefore 
\[
\text{Tr}\left( e^{\alpha \hat A+\left( 1-\alpha \right) \hat B}\right)
<\left( \text{Tr}e^{\hat A}\right) ^\alpha \left( \text{Tr}e^{\hat B}\right)
^{\left( 1-\alpha \right) }, 
\]
when $\hat A\neq \hat B+const.$

Setting 
\[
\hat A=-\beta \left( \hat H+J\left( 1\right) \hat Q\left( 1\right) \right) 
\]
and 
\[
\hat B=-\beta \left( \hat H+J^{\prime }\left( 1\right) \hat Q\left( 1\right)
\right) , 
\]
we can easily obtain 
\[
W\left[ \alpha J+\left( 1-\alpha \right) J^{\prime }\right] >\alpha W\left[
J\right] +\left( 1-\alpha \right) W\left[ J^{\prime }\right] , 
\]
when $J\neq J^{\prime }.$ $QED.$

{\bf Corollary 1:} {\it The map }${\cal J}\rightarrow {\cal Q}${\it \ is
one-to-one.}

{\bf Proof: }Consider the functional

\[
\Lambda \left[ J\right] =W\left[ J\right] -J\left( 1^{\prime }\right)
Q\left( 1^{\prime }\right) {\bf ,} 
\]
where $J$ and $Q$ are considered to be independent. Since $W\left[ J\right] $
is strictly concave $\left( \text{Theorem 1}\right) ,$ it follows that $%
\Lambda \left[ J\right] $ is strictly concave. Therefore if $\Lambda \left[
J\right] $ has an extremum it is unique. Hence if the equation 
\[
\frac{\delta \Lambda \left[ J\right] }{\delta J\left( 1\right) }=\frac{%
\delta W\left[ J\right] }{\delta J\left( 1\right) }-Q\left( 1\right) =0 
\]
has a solution, it is unique. $QED.$

{\bf Proposition 1: }{\it The effective action functional }$\Gamma \left[
Q\right] ${\it \ defined on the set }${\cal Q}${\it \ is strictly convex.}

{\bf Proof:} Consider the family of the functionals $\left\{ \Lambda \left[
J,Q\right] :J\in {\cal J},Q\in {\cal Q}\right\} ,$ 
\[
\Lambda \left[ J,Q\right] =W\left[ J\right] -J\left( 1^{\prime }\right)
Q\left( 1^{\prime }\right) . 
\]
Here $Q$ is considered to be independent of $J.$ Obviously $\Lambda \left[
J,Q\right] $ is linear in $Q.$ The effective action functional can be
defined as 
\begin{equation}
\Gamma \left[ Q\right] =\sup \left\{ \Lambda \left[ J,Q\right] ,J\in {\cal J}%
\right\} .  \label{eff-action1}
\end{equation}
The above expression represents the most general way to define the effective
action functional. In our case two definitions, Eqn $\left( \ref{G[Q]}%
\right) $ and Eqn $\left( \ref{eff-action1}\right) ,$ are equivalent.
Convexity of $\Gamma \left[ Q\right] $ follows from the fact that it is a
supremum of the family of the linear functionals $\left( \text{in }Q\right) $%
. Because by construction for any element $J\in {\cal J},$ there corresponds
only one $Q\in {\cal Q}$ , the functional $\Gamma \left[ Q\right] $ is also
strictly convex. $QED.$

\subsection{Time-dependent probe}

{\bf Theorem 2: }{\it The operator }$W^{\left( 2\right) }\left( x,x^{\prime
}\right) ${\it \ is strictly negative definite}

{\bf Proof: }We have to prove that there exists no function $f\left( {\bf x}%
,\tau \right) \neq \eta \left( \tau \right) $ such that 
\begin{equation}
\int \int f^{*}\left( x\right) W^{\left( 2\right) }\left( x,x^{\prime
}\right) f\left( x^{\prime }\right) dxdx^{\prime }\geq 0.  \label{fWf}
\end{equation}
The operator $W^{\left( 2\right) }\left( x,x^{\prime }\right) $ actually
represents the negative of the density correlation function in the presence
of the {\it time-independent} source $J\left( {\bf x}\right) $: 
\[
W^{\left( 2\right) }\left( x,x^{\prime }\right) =-\frac{\int D\psi ^{\dagger
}D\psi \widetilde{n}\left( x\right) \widetilde{n}\left( x^{\prime }\right)
e^{-S_J\left[ \psi ^{\dagger },\psi \right] }}{\int D\psi ^{\dagger }D\psi
e^{-S_J\left[ \psi ^{\dagger },\psi \right] }}, 
\]
where 
\[
S_J\left[ \psi ^{\dagger },\psi \right] =S\left[ \psi ^{\dagger },\psi
\right] +\int J\left( {\bf x}\right) n\left( x\right) dx, 
\]
and $\widetilde{n}\left( x\right) $ denotes density fluctuation operator, 
\[
\widetilde{n}\left( x\right) =\psi ^{\dagger }\left( x\right) \psi \left(
x\right) -n\left( x\right) . 
\]
Defining a Fourier transform as, 
\begin{eqnarray*}
W^{\left( 2\right) }\left( {\bf x},{\bf x}^{\prime },i\nu _s\right)
&=&\int_0^\beta W^{\left( 2\right) }\left( x,x^{\prime }\right) e^{i\nu
_s\left( \tau -\tau ^{\prime }\right) }d\left( \tau -\tau ^{\prime }\right) ,%
\hspace{0.35cm} \\
\nu _s &=&2\pi s/\beta ,
\end{eqnarray*}
left hand side of $\left( \ref{fWf}\right) $ transforms into 
\begin{equation}
\sum_{\nu _s}\int \int \left( f\left( {\bf x},\nu _s\right) \right)
^{*}W^{\left( 2\right) }\left( {\bf x},{\bf x}^{\prime },i\nu _s\right)
f\left( {\bf x}^{\prime },\nu _s\right) d{\bf x}d{\bf x}^{\prime }.
\label{fWfv}
\end{equation}
Strict concavity of $W\left[ J\right] $ for time-independent case guarantees
that $\nu _s=0$ term is strictly negative definite. For convenience, we
ignore this term from the sum. Lehman representation for $W^{\left( 2\right)
}\left( {\bf x},{\bf x}^{\prime },i\nu _s\right) $ is given by, 
\begin{eqnarray}
W^{\left( 2\right) }\left( {\bf x},{\bf x}^{\prime },i\nu _s\right)
=e^{W\left[ J\right] } &&\sum_{ml}\frac{e^{-\beta E_m}-e^{-\beta E_l}}{%
E_m-E_l+i\nu _s}\times  \nonumber \\
&&\ \times \left\langle m\left| \widetilde{n}\left( {\bf x}\right) \right|
l\right\rangle \left\langle l\left| \widetilde{n}\left( {\bf x}^{\prime
}\right) \right| m\right\rangle ,
\end{eqnarray}
where 
\begin{eqnarray*}
\left( \hat H+\int J\left( {\bf x}\right) \hat n\left( {\bf x}\right)
\right) |m\rangle &=&E_m|m\rangle , \\
\left( \hat H+\int J\left( {\bf x}\right) \hat n\left( {\bf x}\right)
\right) |l\rangle &=&E_l|l\rangle .
\end{eqnarray*}
Therefore an arbitrary $\left( \nu _s\neq 0\right) $ term in the sum $\left( 
\ref{fWfv}\right) $ can be written as, 
\[
e^{W\left[ J\right] }\sum_{ml}\frac{e^{-\beta E_m}-e^{-\beta E_l}}{%
E_m-E_l+i\nu _s}\left| \left\langle m\left| \int \widetilde{n}\left( {\bf x}%
\right) f\left( {\bf x},\nu _s\right) d{\bf x}\right| l\right\rangle \right|
^2, 
\]
or 
\begin{eqnarray*}
e^{W\left[ J\right] }\sum_{E_m<E_l} &&\left| \left\langle m\left| \int 
\widetilde{n}\left( {\bf x}\right) f\left( {\bf x},\nu _s\right) d{\bf x}%
\right| l\right\rangle \right| ^2\times \\
&&\ \times \left( \frac{e^{-\beta E_m}-e^{-\beta E_l}}{E_m-E_l+i\nu _s}+%
\frac{e^{-\beta E_l}-e^{-\beta E_m}}{E_l-E_m+i\nu _s}\right) .
\end{eqnarray*}
After a simple algebra we obtain 
\begin{eqnarray*}
e^{W\left[ J\right] }\sum_{E_m<E_l}2 &&\left| \left\langle m\left| \int 
\widetilde{n}\left( {\bf x}\right) f\left( {\bf x,}\nu _s\right) d{\bf x}%
\right| l\right\rangle \right| ^2\times \\
&&\ \times \frac{e^{-\beta E_l}-e^{-\beta E_m}}{\left( E_l-E_m\right) ^2+\nu
_s^2}\left( E_l-E_m\right) .
\end{eqnarray*}
Obviously, the above expression cannot be positive. Moreover, it cannot be
zero since this would imply that for all $m\neq l$ 
\[
\left\langle m\left| \int \hat n\left( {\bf x}\right) f\left( {\bf x,}\nu
_s\right) d{\bf x}\right| l\right\rangle =0 
\]
and $\int \hat n\left( {\bf x}\right) f\left( {\bf x,}\nu _s\right) d{\bf x}$
commutes with the Hamiltonian. This is impossible unless $f\left( {\bf x,}%
\nu _s\right) $ is constant independent of ${\bf x}$ or $f\left( {\bf x,}%
\tau \right) $ is a function of $\tau $ only. This case, however, was
excluded from the very beginning. Therefore, the operator $W^{\left(
2\right) }\left( x,x^{\prime }\right) $ is strictly negative definite. $QED.$

{\bf Proposition 2}: {\it The operator }$W^{\left( 2\right) }{\bf (x},{\bf x}%
^{\prime },\omega )=W^{\left( 2\right) }\left( {\bf x},{\bf x}^{\prime
},i\nu _s\right) |_{i\nu _s\rightarrow \omega +i\eta }${\it \ has no zero
eigenvalues when }$\omega ${\it \ is located in the upper half of the
complex plane\ including the real axis.}

{\bf Proof: }Suppose $f\left( {\bf x}\right) $ is an eigenvector
corresponding to a zero eigenvalue then 
\begin{equation}
\int \int f^{*}\left( {\bf x}\right) W^{\left( 2\right) }{\bf (x},{\bf x}%
^{\prime },\omega )f\left( {\bf x}^{\prime }\right) d{\bf x}d{\bf x}^{\prime
}=0.  \label{zerocond}
\end{equation}
Using Lehman representation we obtain 
\begin{eqnarray*}
0=\sum_{E_{m}<E_{l}} &&\left| \left\langle m\left| \int \widetilde{n}\left( 
{\bf x}\right) f\left( {\bf x}\right) d{\bf x}\right| l\right\rangle \right|
^{2}\times \\
&&\times \left( \frac{e^{-\beta E_{m}}-e^{-\beta E_{l}}}{z-\left(
E_{l}-E_{m}\right) }-\frac{e^{-\beta E_{m}}-e^{-\beta E_{l}}}{z+\left(
E_{l}-E_{m}\right) }\right)
\end{eqnarray*}
where $z=\omega +i\eta .$ Consider the imaginary part of the above
expression, 
\begin{eqnarray*}
0=-4Re\left( z\right) Im\left( z\right) &&\sum_{E_{m}<E_{l}}\left|
\left\langle m\left| \int \widetilde{n}\left( {\bf x}\right) f\left( {\bf x}%
\right) d{\bf x}\right| l\right\rangle \right| ^{2}\times \\
&&\times \frac{\left( E_{l}-E_{m}\right) \left( e^{-\beta E_{m}}-e^{-\beta
E_{l}}\right) }{\left| z^{2}-\left( E_{l}-E_{m}\right) ^{2}\right| ^{2}}
\end{eqnarray*}
\[
\Rightarrow Re\left( z\right) Im\left( z\right) =0. 
\]
By Theorem 2, the real part of $z$ cannot be zero, since in this case $%
W^{\left( 2\right) }{\bf (x},{\bf x}^{\prime },\omega =i\nu _{s})$ is
strictly negative definite and condition $\left( \ref{zerocond}\right) $
cannot be satisfied. Therefore we necessarily obtain that $Im\left( z\right)
=0$ and 
\[
\omega +i\eta =%
\mathop{\rm real}
\hspace{0.55cm}\Rightarrow \hspace{0.55cm}\omega =%
\mathop{\rm real}
-i\eta 
\]
Therefore $W^{\left( 2\right) }{\bf (x},{\bf x}^{\prime },\omega )$ may have
zero eigenvalues only when $\omega $ is located in the lower half of the
complex plane\ excluding the real axis{\it .} $QED.$

{\bf Proposition 3:} $\Gamma ^{\left( 2\right) }({\bf x},{\bf y,-}i\nu _s)$
has a unique analytic continuation $\Gamma ^{\left( 2\right) }({\bf x},{\bf %
y,\,}\omega )$ such that

\begin{enumerate}
\item  it does not have any zeros in the upper half of the complex plane $%
\omega $ including the real axis

\item  $\left[ \Gamma ^{\left( 2\right) }({\bf x},{\bf y,\,}\omega )\right]
^{-1}\sim \frac{1}{\left| \omega \right| }$, $\left( \omega \rightarrow
\infty \right) $.
\end{enumerate}

{\bf Proof:}

Existence follows from the fact that $\Gamma ^{\left( 2\right) }({\bf x},%
{\bf y,\,}\omega )$ can be defined as $-\left[ W^{\left( 2\right) }({\bf x},%
{\bf y,\,}\omega )\right] ^{-1}.$ Suppose there exist two analytic
continuations of $\Gamma ^{\left( 2\right) }({\bf x},{\bf y,-}i\nu _{s})$
with the properties 1 and 2. Then $W^{\left( 2\right) }{\bf (x},{\bf x}%
^{\prime },i\nu _{s})$ will have two different analytic continuations which
is impossible. $QED.$


\begin{references}
\bibitem{HK}  P. Hohenberg and W. Kohn, Phys. Rev. {\bf 136B, }864(1964).

\bibitem{SDFT}  U. von Barth, L. Hedin, J. Phys. C{\bf 5}, 1629(1972); M. M.
Pant, A. K. Rajagopal, Sol. State Comm., {\bf 10}, 1157(1972).

\bibitem{CDFT}  G. Vignale, M. Rasolt, Phys. Rev. Lett., {\bf 59},
2360(1987).

\bibitem{fukuda}  R. Fukuda, T. Kotani, and S. Yokojima, Prog. Theor. Phys. 
{\bf 92, }833(1994).

\bibitem{fukuda1}  R. Fukuda, M. Komachiya, S. Yokojima, Y. Suzuki, K.
Okumura and T. Inagaki, Prog. Theor. Phys. Suppl. {\bf 121}, (1995).

\bibitem{Mermin}  N. D. Mermin, Phys. Rev., {\bf 137}, A1414(1965).

\bibitem{KS}  W. Kohn and L. J. Sham, Phys. Rev. {\bf 140, }A1133(1965).

\bibitem{HKS}  P. C. Hohenberg, W. Kohn, L.J. Sham, Adv. Quant. Chem. {\bf %
21,} 7(1990).

\bibitem{DFTbook1}  R. M. Dreizler, E. K. U. Gross, {\it Density-Functional
Theory}, (Springer-Verlag, Berlin, Heidelberg, 1990) and references therein.

\bibitem{DFTbook2}  R. G. Parr, W. Yang, {\it Density-Functional Theory of
Atoms and Molecules}, (Oxford University Press, New York, 1989).

\bibitem{okumura}  K. Okumura, Int. J. Mod. Phys. {\bf 11,} 65(1996).

\bibitem{sham}  L. J. Sham, Phys. Rev. {\bf B32,} 3876(1985).

\bibitem{OEP}  J. B. Krieger and Yan Li, G. J. Iafrate, Phys. Rev. {\bf A45, 
}101(1992).

\bibitem{valiev}  M. Valiev and G. W. Fernando, Phys. Lett. A {\bf 227},
265(1997).

\bibitem{SK}  L. J. Sham and W. Kohn, Phys. Rev. {\bf 145,} 561(1966).

\bibitem{GW}  M. Schl\"uter, L.J. Sham, Adv. Quant. Chem. {\bf 21,}
97(1990), and references therein.

\bibitem{GW1}  R. Del Sole, L. Reining, R. W. Godby, Phys. Rev. {\bf B49,}
8024(1994), and references therein.

\bibitem{peter}  M. Petersilka, U.J. Gossmann, E.K.U. Gross, Phys. Rev.
Lett. {\bf 76,} 1212(1996).

\bibitem{excitation1}  R. Bauernschmitt, R. Ahlrichs, Chem. Phys. Lett. {\bf %
256}, 454(1996).

\bibitem{GH}  E. K. U. Gross, W. Kohn, Adv.Quant. Chem. {\bf 21,} 255(1990).

\bibitem{legendre1}  the usefulness of the Legendre transformation in
density-functional theory was earlier recognized by E.H. Lieb in {\em %
Density Functional Methods in Physics}, edited by R.M. Dreizler and J. da
Providencia (Plenum Press, New York,1985); and Nalewajski, J. Chem. Phys. 
{\bf 78}, 6112(1983) and references therein.

\bibitem{valiev1}  M. Valiev and G. W. Fernando, Phys. Rev. B{\bf 54},
7765(1996).

\bibitem{negele}  J. W. Negele and H. Orland, {\it Quantum Many-Particle
Systems}, (Addison-Wesley, 1995).

\bibitem{Levy}  A. G\"orling, M. Levy, Phys. Rev. {\bf B47, }13105(1993).

\bibitem{Levy1}  A. G\"orling, M. Levy, Phys. Rev. {\bf A50, }196(1994).

\bibitem{ryder}  L. H. Ryder, {\it Quantum Field Theory} (Cambridge
University Press, New York, 1985).

\bibitem{rao}  S. S. Rao, {\it Optimization theory and applications},
(Halsted Press, New York, 1978).

\bibitem{justin}  J. Zinn-Justin, {\it Quantum Field Theory and Critical
Phenomena},{\it \ }(Clarendon Press, Oxford, 1993).

\bibitem{israel}  some of the techniques used in the proof were borrowed
from R. B. Israel, {\it Convexity in the Theory of Lattice Gases},
(Princeton University Press, Princeton, New Jersey, 1979).

\bibitem{holder}  A. E. Taylor, {\it General Theory of Functions and
Integration}, (Dover Publications, Inc., New York, 1985).
\end{references}
\end{document}